\documentclass[11pt,a4paper]{article}
\pdfoutput=1
\usepackage{jcappub}
\usepackage[utf8]{inputenc}

\usepackage{caption}
\usepackage[usenames,dvipsnames,svgnames,table]{xcolor}
\usepackage{amsmath}
\usepackage{gensymb}
\usepackage{mathrsfs}
\usepackage{slashed}
\graphicspath{{Figures/}}
\usepackage{float}
\usepackage{color}
\usepackage{physics}
\usepackage{amsbsy}
\usepackage{subcaption}
\usepackage{caption}
\usepackage{hyperref}
\hypersetup{
    colorlinks=true,
    linkcolor=blue,
    filecolor=magenta,      
    urlcolor=blue,
}

\newcommand*\Bell{\ensuremath{\boldsymbol\ell}}

\usepackage[T1]{fontenc}

\notoc

\begin{document}

\title{Taller in the saddle: constraining CMB physics using saddle points}
\author[a,b,1]{Dylan L. Jow,\note{Corresponding author.}}
\author[a,c,d]{Dagoberto Contreras,}
\author[a]{Douglas Scott,}
\author[e]{and Emory F. Bunn}

\affiliation[a]{Department of Physics and Astronomy, University of British Columbia, \\6224 Agricultural Road, Vancouver, BC, Canada}
\affiliation[b]{Department of Physics, University of Toronto, ON M5S 1A7, Canada}
\affiliation[c]{Department of Physics and Astronomy, York University, ON M3J 1P3, Canada}
\affiliation[d]{Perimeter Institute, 31 Caroline St N, Ontario, Canada}
\affiliation[e]{Physics Department, University of Richmond, Richmond, VA 23173, USA}

\emailAdd{djow@physics.utoronto.ca}

\abstract{The statistics of extremal points in the cosmic microwave background (CMB) temperature (hot and cold spots) have been well explored in the literature, and have been used to constrain models of the early Universe. Here, we extend the study of critical points in the CMB to the set that remains after removing extrema, namely the saddle points. We perform stacks of temperature and polarization about temperature saddle points in simulations of the CMB, as well as in data from the \textit{Planck} satellite. We then compute the theoretical profile of saddle-point stacks, given the underlying power spectra of the CMB. As an example of the utility of such stacks, we constrain models of cosmic birefringence, and compare the constraining power of the saddle points with that of extremal points. We find that, in the specific example of birefringence, we can place tighter constraints using saddle points in our analysis than using extrema. In fact, we find our saddle-point analysis yields close to optimal constraints, as seen by comparing to a power spectrum analysis. We, therefore, suggest that stacking on saddle points may, in general, be a useful way of testing for non-standard physics effects that change the CMB power spectra.}


\maketitle
\hfil
\flushbottom
\section{Introduction}

Essential to the process of extracting cosmological information from the cosmic microwave background (CMB) radiation is a full characterization of its statistics. The temperature fluctuations in the CMB are consistent with an isotropic Gaussian random field, with some minor exceptions (e.g. due to gravitational lensing, and the integrated Sachs-Wolfe effect)  \citep{IandS2015}. Thus, to appreciate the CMB we must understand the statistics of a 2D Gaussian random field on a 2-sphere (i.e.\,the sky). 

\begin{figure}[htb]
\centering
\includegraphics[width=0.8\columnwidth]{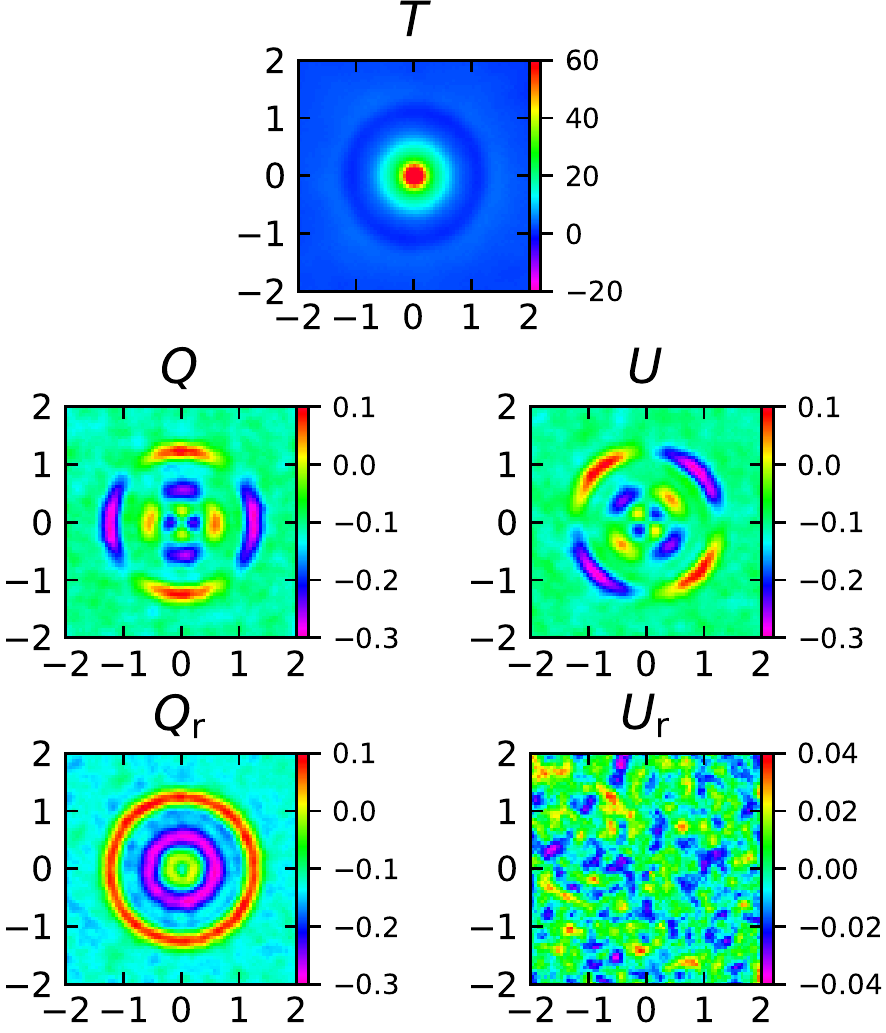}
\caption{Un-oriented stacks of $T$, $Q$, $U$, $Q_\mathrm{r}$, and $U_\mathrm{r}$ around temperature maxima for the \textit{Planck} \texttt{SMICA} CMB map. The stacks are in units of $\mu$K, and the $x$- and $y$-axes are in units of degrees. See Appendix~\ref{sec:peak_stacks} for more details.}
\label{fig:hsp_SMICA}
\end{figure}

The study of 1D Gaussian random fields originated in the study of telecommunications \citep{Rice1944,Rice1945}. This approach was adapted for cosmological applications in the seminal 1986 paper by Bardeen, Bond, Kaiser \& Szalay, in which the perturbations of the evolving early Universe were modelled as a 3D Gaussian field \citep{BBKS}. These results were particularized to the 2D case of Gaussian fields on a sphere by Bond \& Efstathiou in 1987 \citep{BondEfs1987}. The focus of these papers was the statistical properties of extremal points, i.e.\,local maxima and minima in Gaussian random fields, which has led to a fruitful avenue of study of the statistics of collapsing peaks in large-scale structure. The average profiles of the temperature and polarization fields around temperature peaks were calculated in the flat-sky limit by the \textit{WMAP} team \citep{Komatsu2011}, and a more general formalism for calculating these profiles, valid at large scales (i.e.\,with no flat-sky approximation) and allowing for biased peak eccentricity, was developed by Marcos-Caballero et al.~\citep{MC2016}. Stacks of temperature and polarization patches around temperature peaks in CMB data can be compared to these calculations to test cosmological models. Importantly, the theoretical predictions for the profiles of these stacks rely on the assumptions of Gaussianity and statistical isotropy, and, therefore, they can be used as general tests for these assumptions. Using such stacks, the \textit{WMAP}~\citep{Komatsu2011} and \textit{Planck}~\citep{IandS2015} teams have found complete consistency with the standard model. Moreover, the bullseye-like plots produced by both the \textit{WMAP} and \textit{Planck} teams have been frequently used as a simple real-space way to visualize the oscillatory physics of the model \citep{Komatsu2011,IandS2015,2014A&A...571A...1P,2013ApJS..208...19H}. In order to contrast with our new stacks, we show the standard averages around temperature extrema in Figure~\ref{fig:hsp_SMICA}. The fact that these stacks match the predictions has been used to constrain non-standard physics such as cosmic birefringence \citep{2016A&A...596A.110P}.

As well as maxima and minima, there is a third type of critical point in 2D: saddle points. Though the formalism developed in the literature is valid for general critical points, the focus has been almost entirely on extrema, as opposed to saddle points. However, since polarization is sourced by quadrupole anisotropies in temperature \citep{1997NewA....2..323H}, temperature saddle points, which take the form of quadrupoles on the sky, are natural objects to study. Moreover, scalar, vector, and tensor perturbations source specific modes of quadrupole anisotropies \citep{1997NewA....2..323H}, and so saddle points may be of interest in studying the characteristics of different sources of perturbation. Some of the general statistical properties of saddle points in Gaussian random fields have already been studied (see e.g. Refs.~\citep{1998ApJ...507...31N,2003MNRAS.344..905D,2005PhRvD..72d3004H}); however, the use of stacks to visualize and constrain physics has been primarily focused on extrema. 

In Section~\ref{sec:stacks} of this paper, we present stacks around temperature saddle points for simulated CMB maps containing purely scalar, vector, or tensor perturbations, as well as real \textit{Planck} data. In Section~\ref{sec:saddle_stats}, we compute the expected profiles of temperature and polarization about temperature saddle points in the CMB in the flat-sky limit. For completeness, we also present the result computed for the spherical sky, extending the formalism of earlier papers, particularly Ref.~\citep{MC2016}. We end with an example of the application of the use of stacked saddle-point plots for constraining exotic physics, in Section~\ref{sec:biref}. This enables us to explore how the constraining power of saddle points compares with that of extrema.

\section{Data and simulations}
\label{sec:data}

In this work, we use the 2018 full-mission and half-mission \textit{Planck} data \citep{2018arXiv180706205P,2018arXiv180706206P,2018arXiv180706207P}, specifically adopting the \texttt{SMICA} component-separation \citep{2018arXiv180706208P} procedure to obtain maps of the Stokes parameters $T$, $Q$, and $U$. The maps are provided at a resolution of \texttt{HEALPix}\footnote{See \url{http://healpix.sourceforge.net} for more details.}\,$N_\mathrm{side} = 2048$, to which we apply a $10^\prime$ beam. Note that the \textit{Planck} satellite has a $5^\prime$ beam; however, for the purposes of stacking around extrema and saddle points, applying a $10^\prime$ beam produces stacks with less noise, but does not wash out any of the relevant features. We also use simulations of the CMB, with power spectra generated using \texttt{CAMB} \citep{Lewis:1999bs} and the map manipulation tools of \texttt{HEALPix} \citep{Healpix}. We generate three sets of simulations for purely scalar, vector, and tensor perturbations, described in section~\ref{sec:stacks}. For all simulations, we input a fiducial $\Lambda$CDM model into \texttt{CAMB}, except for simplicity we set the number of massive neutrinos to 0 (since massive neutrino models are incompatible with vector modes in \texttt{CAMB}). We generate simulated maps at a resolution of $N_\mathrm{side}=1024$, also smoothed with a $10^\prime$ beam to match the real data. The maps are analysed using \texttt{HEALPy}, the Python wrapper for \texttt{HEALPix} \citep{Healpix}.

\section{Stacks on temperature saddle points}
\label{sec:stacks}

\subsection{Oriented stacking of temperature and polarization}
\label{sec:ortstck}

We are interested in the average profiles of temperature and polarization around temperature saddle points on the sky. For maps of the CMB, we can compute the mean profile around saddle points by averaging over (i.e.\,stacking) small patches of the sky centred around each saddle point. Saddle points are defined as critical points that are neither maxima nor minima. That is, they are points with vanishing first derivatives, and with $\det H \leq 0$, where $H$ is the matrix of second derivatives, the Hessian. In practice, to find the location of temperature saddle points on the sky, we compute the magnitude of the gradient of the temperature, $|\nabla T|^2 = \partial^2_{\theta} T + \frac{1}{\sin^2\theta}\partial^2_{\phi} T$, in the usual spherical coordinates. Critical points coincide with zeros in $| \nabla T|^2$, for which we use minima as proxies, where minima in the gradient are identified by comparing the value of each pixel with the value of its eight nearest neighbours in the \texttt{HEALPix} scheme. We consider minima of $| \nabla T|^2$ with Hessians of negative determinant to be saddle points. Note that this procedure of identifying saddle points will, in general, include points that are not genuinely saddle points due to pixelization issues and the fact that not all minima of $| \nabla T|^2$ are zeros. However, tests show that these additional points give a negligible contribution to the total number of points selected.

Since we are interested in the quadrupolar profile of the saddle points, we need to perform oriented stacking, similar to what is described in Section 8 of Ref.~\citep{IandS2015}. That is, before stacking the patches around each saddle point, we orient them so that the major axes of the saddle points are aligned. This can be achieved by rotating each patch into a local basis where the Hessian of the temperature field at the saddle point is diagonal. For a general field $f$ on a sphere, the Hessian, in the usual spherical coordinate basis $(\bf{e}_\theta, \bf{e}_\phi)$, is given by \citep[e.g.][]{2007AmJPh..75..116M}
\begin{equation}
H[f] =
\begin{bmatrix}
	\partial_{\theta\theta} f & \frac{1}{\sin\theta} \partial_{\phi\theta} f - \frac{\cos\theta}{\sin^2\theta}\partial_\phi f \\
	\frac{1}{\sin\theta} \partial_{\phi\theta} f - \frac{\cos\theta}{\sin^2\theta}\partial_\phi f & \frac{1}{\sin^2\theta}\partial_{\phi\phi}f + \frac{\cos\theta}{\sin\theta} \partial_\theta f
\end{bmatrix}.
\end{equation}
Thus, when stacking on saddle points, we orient the patches so that $H[T]$ is diagonal at the centre. To ensure that the patches are consistently oriented, we always select a right-handed basis, where the $x$-axis is chosen so that $(H_{xx} - H_{yy}) \geq 0$. In order to compute the derivatives of a full-sky map, we use the \texttt{HEALPy} function \texttt{alm2map\_der1}, which computes the derivatives from the spherical harmonic transform of a given map. Higher derivatives are computed using successive applications of this function.

We stack both the temperature and polarization fields on the locations of temperature saddle points. Polarization of the light from the CMB is measured using the Stokes parameters $Q$ and $U$, where positive and negative $Q$ describe the degree of linear polarization in the $\mathbf{e}_\theta$ and $\mathbf{e}_\phi$ directions, respectively, and positive and negative $U$ describe the polarization in the $(\mathbf{e}_\theta \pm \mathbf{e}_\phi)/\sqrt{2}$ directions, respectively. There is a third Stokes parameter, $V$, which describes the degree of circular polarization; however, since CMB polarization is generated through Thomson scattering, which does not produce circular polarization, $V$ is not sourced and hence is typically ignored \citep{1997NewA....2..323H}. The polarization pseudo-vector can be computed from the Stokes parameters by
\begin{subequations}
\begin{equation}
P = \sqrt{Q^2+U^2},
\end{equation}
\begin{equation}
\psi_\mathrm{pol} = \frac{1}{2}\arctan{\frac{U}{Q}},
\end{equation}
\end{subequations}
where $P$ is the magnitude of the polarization, and $\psi_\mathrm{pol}$ is the angle of the polarization pseudo-vector measured from $\mathbf{e}_\theta$ \citep{IandS2015}. Polarization is a spin-2 quantity and transforms with rotations of the local basis according to 
\begin{equation}
Q' + iU' = (Q + iU) e^{-2i\beta},
\label{eqn:polrot}
\end{equation}
where $\beta$ is the rotation angle. Thus, when stacking we both project $(Q+iU)$ onto the new basis and multiply by a phase to obtain the oriented patch. The stacks are given in flat-sky coordinates
\begin{equation}
x = \varpi \cos\psi, \quad y = \varpi \sin\psi,
\end{equation}
where $\varpi$ is the angular distance from the origin, and $\psi$ is the angle measured from the $x$-axis. Additionally, we can obtain the so-called radial Stokes parameters $Q_\mathrm{r}$ and $U_\mathrm{r}$ from 
\begin{equation}
Q_\mathrm{r} + iU_\mathrm{r} = (Q + iU) e^{-2i\psi} .
\end{equation}
Here, $Q_\mathrm{r}$ describes the radial component of the polarization and is positive or negative for radial or tangential polarization, respectively. As with $Q$ and $U$, $U_\mathrm{r}$ describes the polarization pattern of $Q_\mathrm{r}$ rotated by $45^\circ$ \citep{1997PhRvD..55.7368K}.

It is often convenient to expand polarization using the spin-2 spherical harmonics, $_{\pm2} Y^m_\ell$, as \citep{1997PhRvD..55.7368K, 1997NewA....2..323H}
\begin{equation}
(Q \pm iU)(\theta,\phi) = \sum^{\infty}_{\ell=2} \sum^\ell_{m=-\ell} a^{\pm2}_{\ell m} {}_{\pm2}Y^m_\ell (\theta,\phi).
\end{equation}
We use the expansion coefficients to define 
\begin{subequations}
\begin{align}
a^{E}_{\ell m} & \equiv \frac{1}{2}\big(a^{2}_{\ell m} + a^{-2}_{\ell m}\big), \\
a^{B}_{\ell m} & \equiv \frac{-i}{2}\big(a^{2}_{\ell m} - a^{-2}_{\ell m}\big).
\end{align}
\end{subequations}
These can be used as coefficients in a spin-0 spherical harmonic expansion to obtain rotationally invariant quantities $E$ and $B$, which describe the curl-less and divergence-less parts of the polarization pattern, respectively \citep{1997NewA....2..323H}.

\subsection{Saddle-point stacks for scalar, vector, and tensor maps}
\label{sec:svt_stacks}

\begin{figure}[htb]
\centering
\includegraphics[width=\columnwidth]{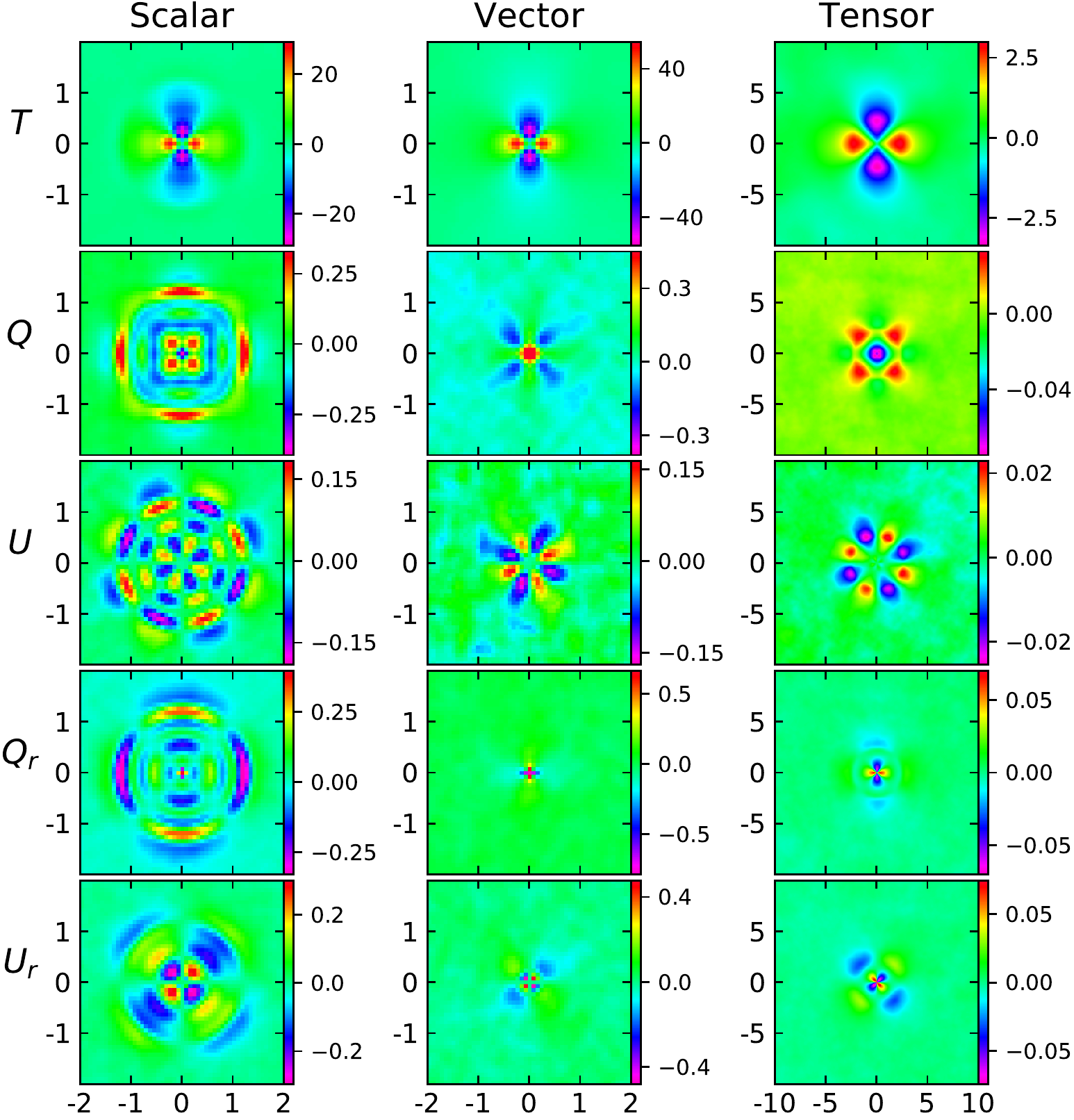}
\caption{Oriented stacks of $T$, $Q$, $U$, $Q_\mathrm{r}$, and $U_\mathrm{r}$ (rows) on temperature saddle points for simulated scalar, vector, and tensor maps (columns), in units of $\mu$K. The $x$- and $y$-axes label the flat-sky coordinates, given in degrees. Note that the tensor stacks are $20^\circ \times 20^\circ$, while the scalar and vector stacks are $4^\circ \times 4^\circ$.}
\label{fig:TQUQrUrxsvt}
\end{figure}

Anisotropies in the CMB are sourced by gravitational redshift induced by perturbations, $h_{\mu\nu}$, to the metric,
\begin{equation}
g_{\mu\nu} = a^2(\eta_{\mu\nu} + h_{\mu\nu}),
\end{equation}
where $\eta_{\mu\nu}$ is the background Minkowski metric \citep{2003moco.book.....D}. Any general symmetric tensor, such as the metric perturbation, can be separated into three mutually independent components that transform as scalars, vectors, and tensors, respectively, under transformations of the Poincar\'{e} group \citep{1986ApJ...308..546A,1997NewA....2..323H}. Thus, CMB anisotropies are sourced by three distinct modes of perturbation. As discussed in Section~\ref{sec:data}, we simulate maps of the CMB for purely scalar, vector, and tensor perturbations, separately. We then compute the temperature saddle-point stacks for each of these simulated maps. Figure~\ref{fig:TQUQrUrxsvt} shows a side-by-side comparison of the stacks of $T$, $Q$, $U$, $Q_\mathrm{r}$, and $U_\mathrm{r}$ around temperature saddle points for each of the scalar, vector, and tensor maps. Immediately we see that the temperature stacks are simply quadrupoles about the centre, as expected. Since tensor perturbations manifest themselves on a larger scale than scalar and vector perturbations \citep{1997NewA....2..323H}, the features in the tensor stacks extend to larger scales. For this reason, while the scalar and vector stacks are shown for $4^\circ \times 4^\circ$ patches of the sky, the tensor stacks are shown for $20^\circ \times 20^\circ$ patches.

In order to extract the salient information contained in Figure~\ref{fig:TQUQrUrxsvt}, we compute the Fourier-Bessel expansion of the stacks. Any 2D function, $f(r,\phi)$, within a disk of radius $a$ with Dirichlet boundary conditions (i.e. $f(a,\phi)=0$) can be expanded according to
\begin{subequations}
\begin{equation}
f(r,\phi) = \sum_{n=1}^\infty \sum_{m=-\infty}^\infty P_{nm} \Psi_{nm}(r,\phi),
\label{eqn:FB_expansion}
\end{equation}
where
\begin{equation}
\Psi_{nm} = \frac{1}{\sqrt{2\pi N_{nm}}} J_m \Big(\frac{x_{mn}}{s} r\Big) e^{im\phi},
\label{eqn:FB_efunc}
\end{equation}
and
\begin{equation}
N_{nm} = \frac{s^2}{2} J^2_{m+1}(x_{mn}).
\label{eqn:FB_norm}
\end{equation}
\end{subequations}
Here, $J_m$ is the Bessel function of the first kind of order $m$, and $x_{mn}$ is the $n$th zero of $J_m$. The $\Psi_{nm}$ form a complete orthonormal basis on the disk $r\leq s$, and we can compute the expansion coefficients as
\begin{equation}
P_{nm} = \int^s_0 \int^{2\pi}_0 f(r,\phi) \Psi^{*}(r,\phi) rdrd\phi.
\end{equation}
The Fourier-Bessel expansion is the flat analogue to the spherical-harmonic transformation on a sphere. Note that this expansion implicitly assumes Dirichlet boundary conditions at some radius $r=s$. In the case of CMB saddle-point stacks, this generally holds, since both the temperature and polarization stacks vanish at infinity, and in all cases we compute the stacks for a radius sufficiently large that the stacks are effectively zero at the boundary. For the scalar and vector stacks, this radius is $s = 2^\circ$, while it is $s = 10^\circ$ for the tensor stacks. Also note that when computing the Fourier-Bessel expansion of the stacks we treat the independent Stokes parameters as single complex quantities. From the coefficients, we compute their averages over $n$ and $m$ ($\langle |P_{mn} |^2 \rangle_n$ and $\langle |P_{mn} |^2 \rangle_m$, respectively) to determine the relative presence of various modes. 

\begin{figure}[htb]
\centering
\includegraphics[width=0.45\columnwidth]{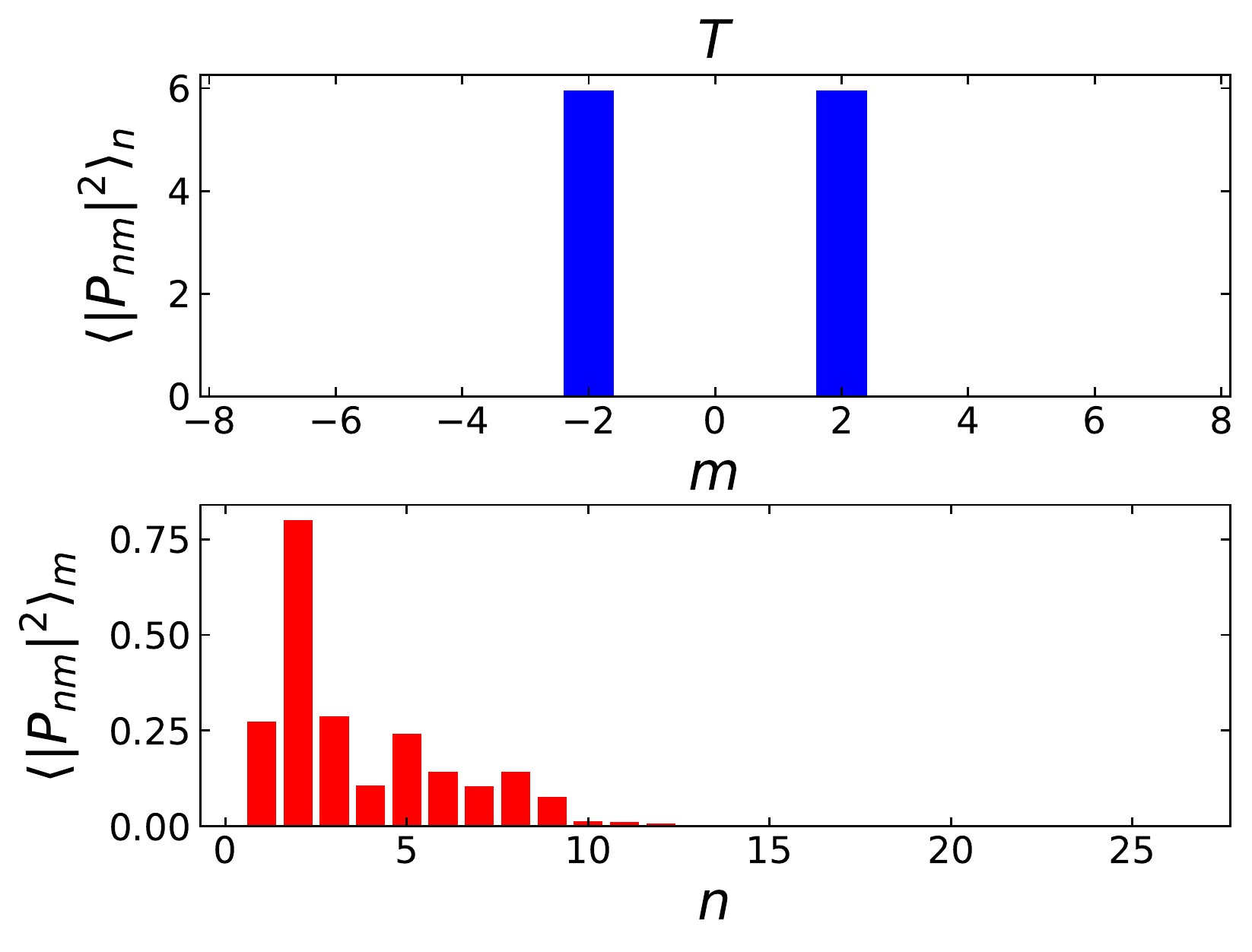}
\medskip

\includegraphics[width=0.47\columnwidth]{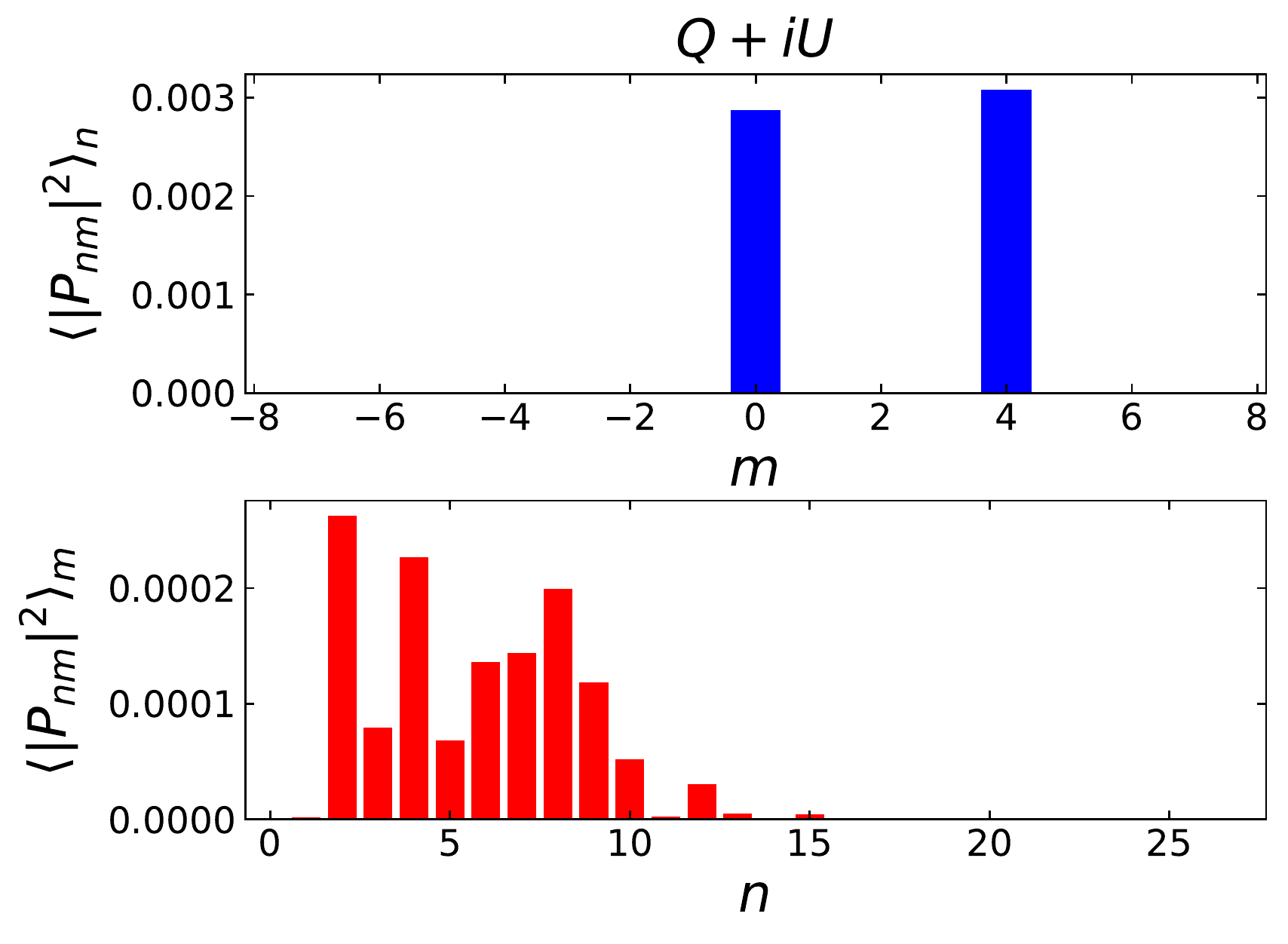}
\includegraphics[width=0.47\columnwidth]{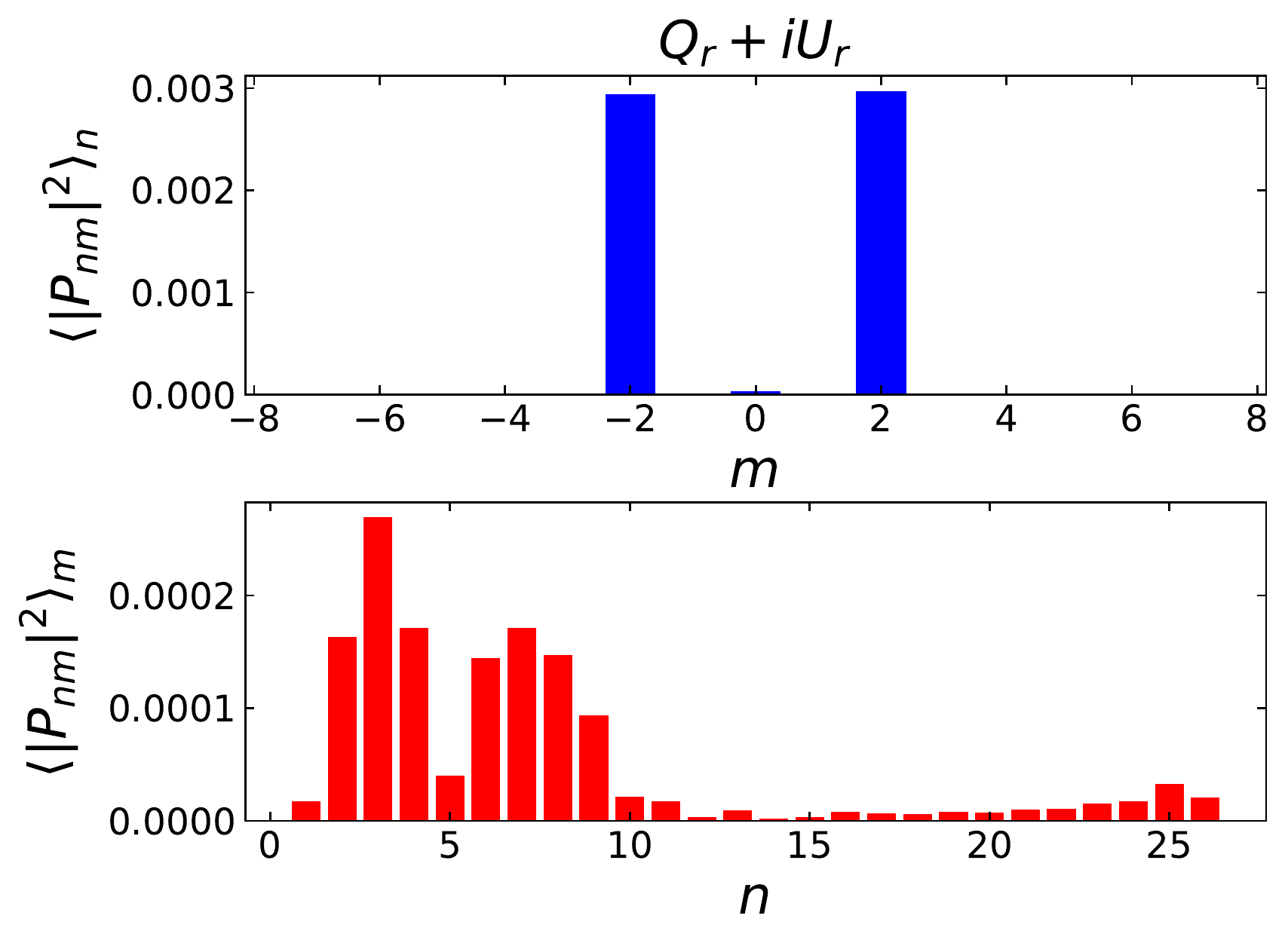}
\caption{$\langle |P_{mn} |^2 \rangle_n$ and $\langle |P_{mn} |^2 \rangle_m$ computed for the $T$, $(Q+iU)$, and $(Q_\mathrm{r} + iU_\mathrm{r})$ stacks around temperature saddle points for the simulated scalar CMB. When computing the Fourier-Bessel expansion of the stacks we treat the independent Stokes parameters as single complex quantities. As expected, the $T$ and $(Q_\mathrm{r} + iU_\mathrm{r})$ stacks have only a quadrupole ($m=\pm2$) moment. Transforming $(Q_\mathrm{r}+iU_\mathrm{r})$ to $(Q+iU)$, by multiplying by $e^{2i\psi}$, shifts the modes up by $2$, yielding an $m=0$ and $m=4$ combination in $(Q+iU)$.}
\label{fig:CmCn_scal}
\end{figure}

Figure~\ref{fig:CmCn_scal} shows the averaged expansion coefficients for the stacks of the simulated scalar map. As we discuss further in Section~\ref{sec:saddle_stats}, the $T$ and $(Q_\mathrm{r} +iU_\mathrm{r})$ stacks contain only $m=\pm2$ modes; i.e.\,they are purely quadrupolar. When transforming from $(Q_\mathrm{r}+iU_\mathrm{r})$ to $(Q+iU)$, we multiply by $e^{2i\psi}$. Thus, the $m=\pm2$ modes in $(Q_\mathrm{r}+iU_\mathrm{r})$ are shifted up by 2 when going to $(Q+iU)$, yielding the observed $m=0$ and $m=4$ modes. We can then decompose $(Q+iU)$ into two distinct types of mode. Since the Fourier component for $m=0$ is real, when splitting $(Q+iU)$ into real and imaginary parts to obtain $Q$ and $U$, we will find that $Q$ is a combination of $m=0$ and $4$ modes, while $U$ is purely $m=4$. Thus, the $T$, $U$, $Q_\mathrm{r}$, and $U_\mathrm{r}$ stacks are all purely one mode, whereas $Q$ also has a monopole contribution. We have checked that this holds true for both the vector and tensor maps as well.

\begin{figure}[htb]
\centering
\includegraphics[width=0.46\columnwidth,height=0.404\columnwidth]{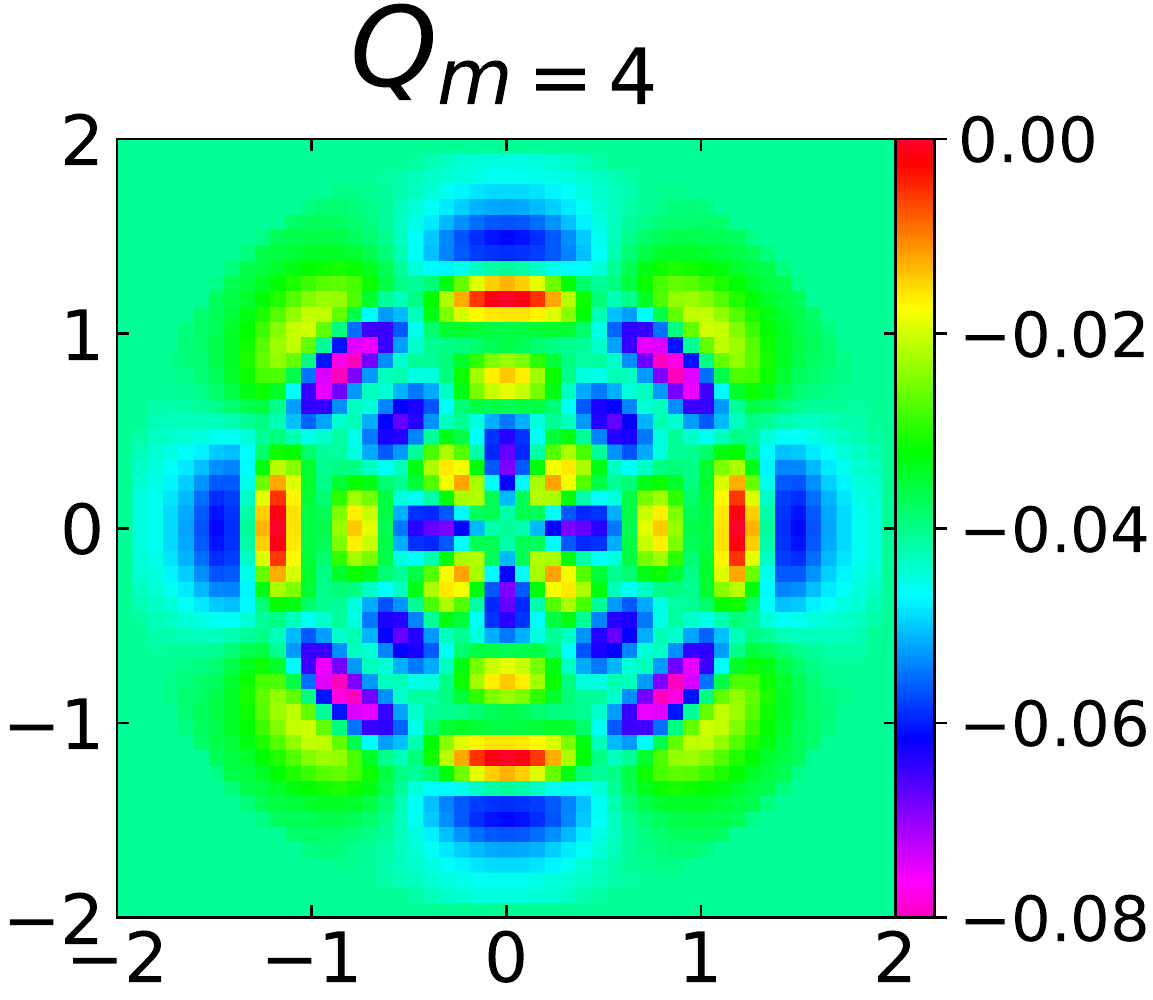}
\includegraphics[width=0.45\columnwidth]{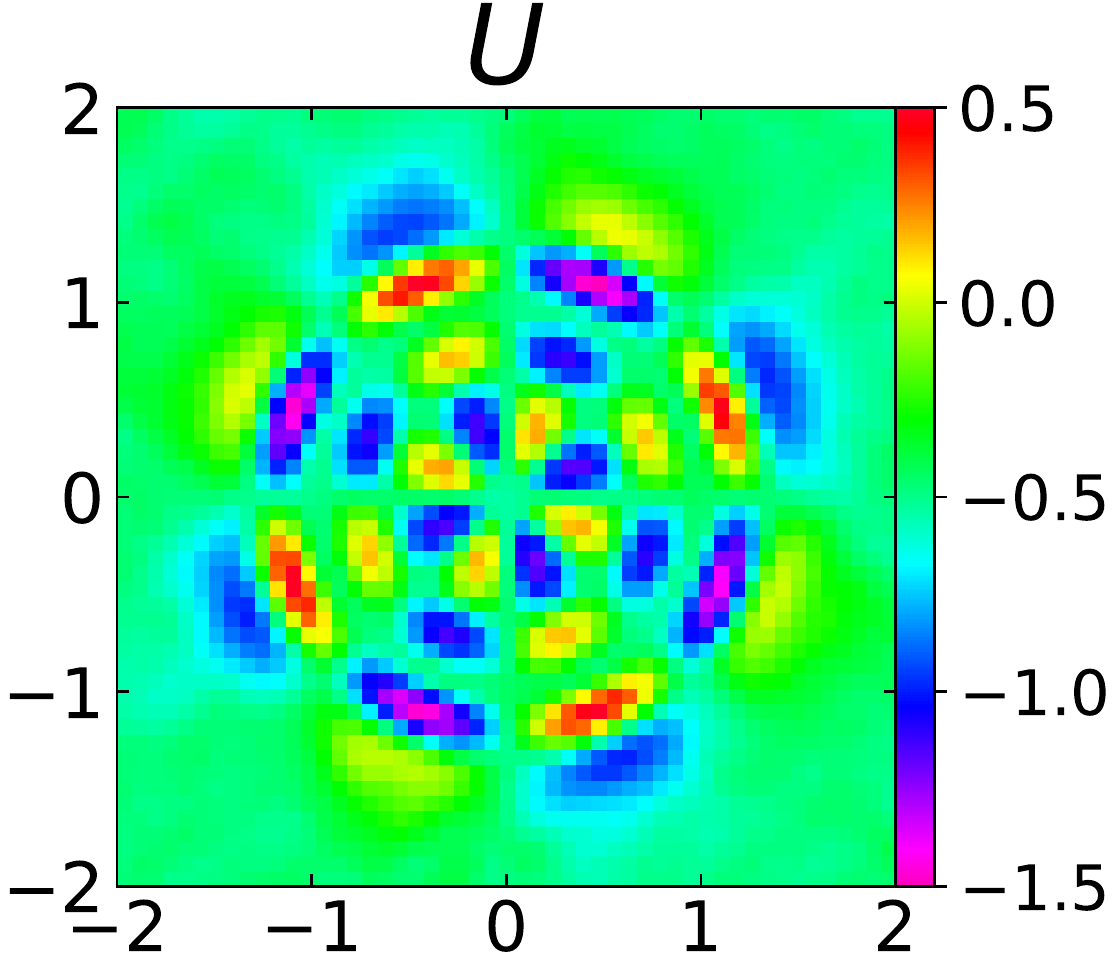}
\caption{Comparison between the full $U$ stack and the $m=4$ part of the $Q$ stack on temperature saddle points for the simulated scalar maps. Note that by removing the $m=0$ mode, the $Q$ and $U$ stacks become identical, but rotated by $22.5^\circ$. We find that this also holds true for both the vector and tensor stacks.}
\label{fig:Q4_U}
\end{figure}

\begin{figure}[htb]
\centering
\includegraphics[width=0.45\columnwidth]{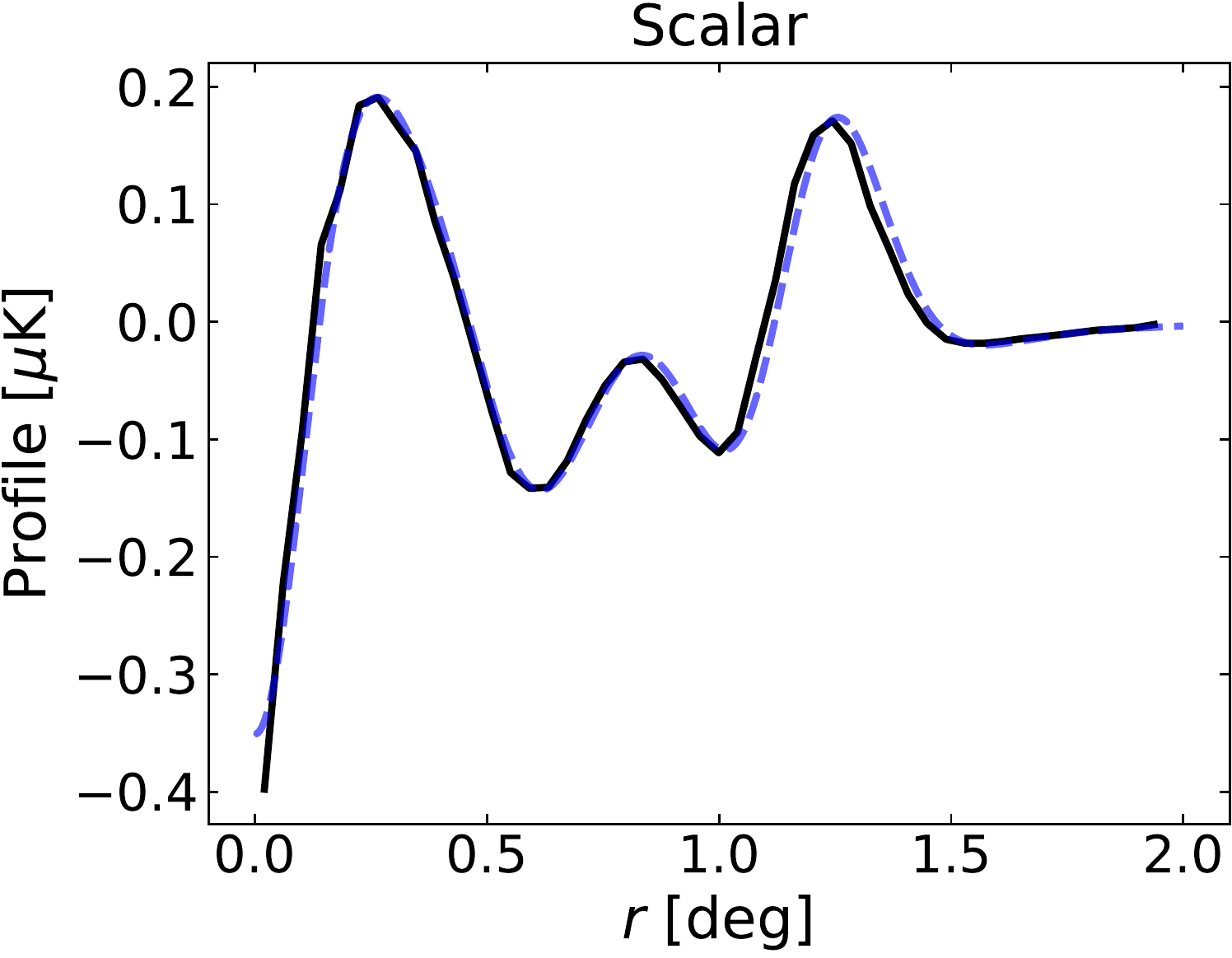}
\includegraphics[width=0.44\columnwidth]{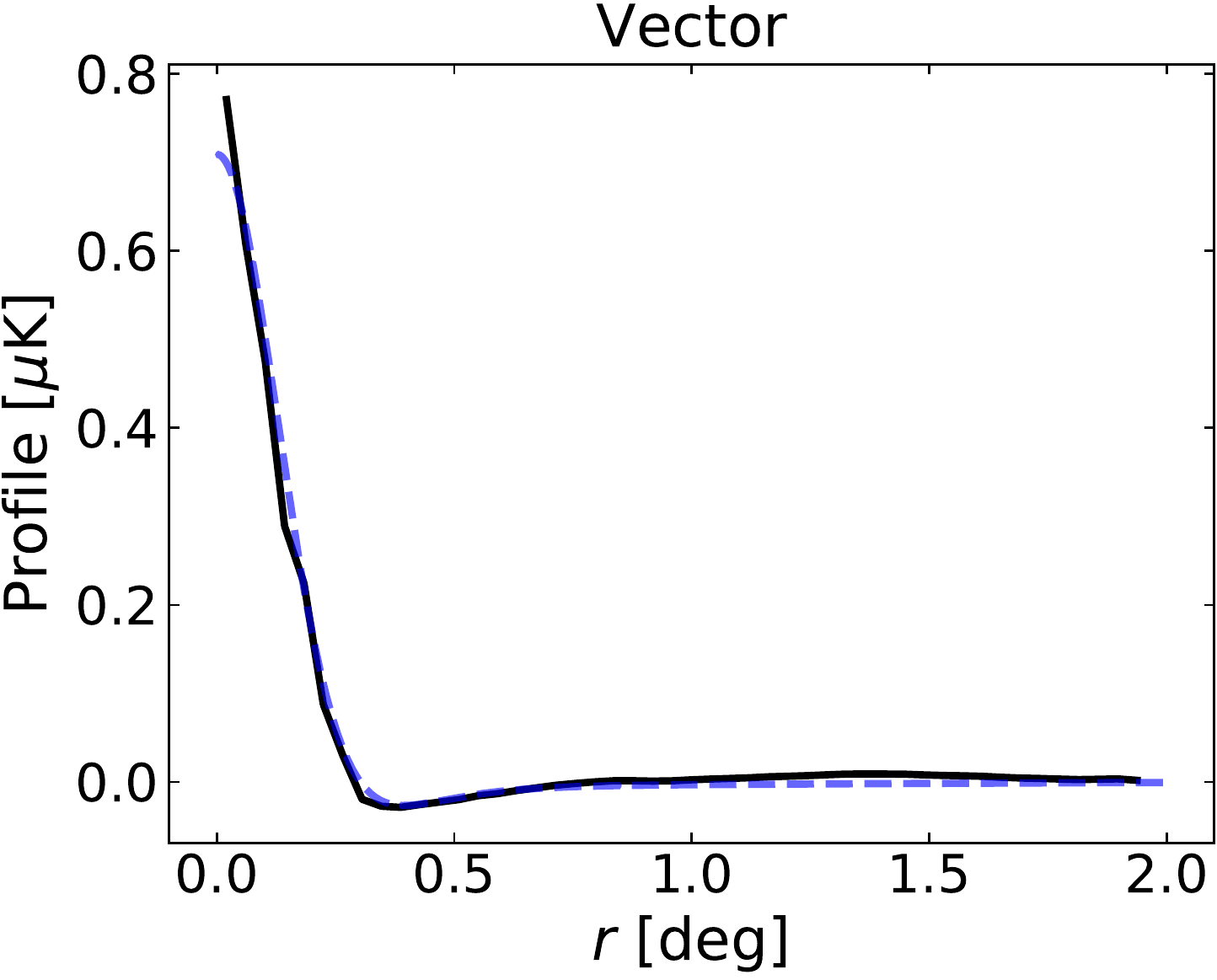}
\includegraphics[width=0.46\columnwidth]{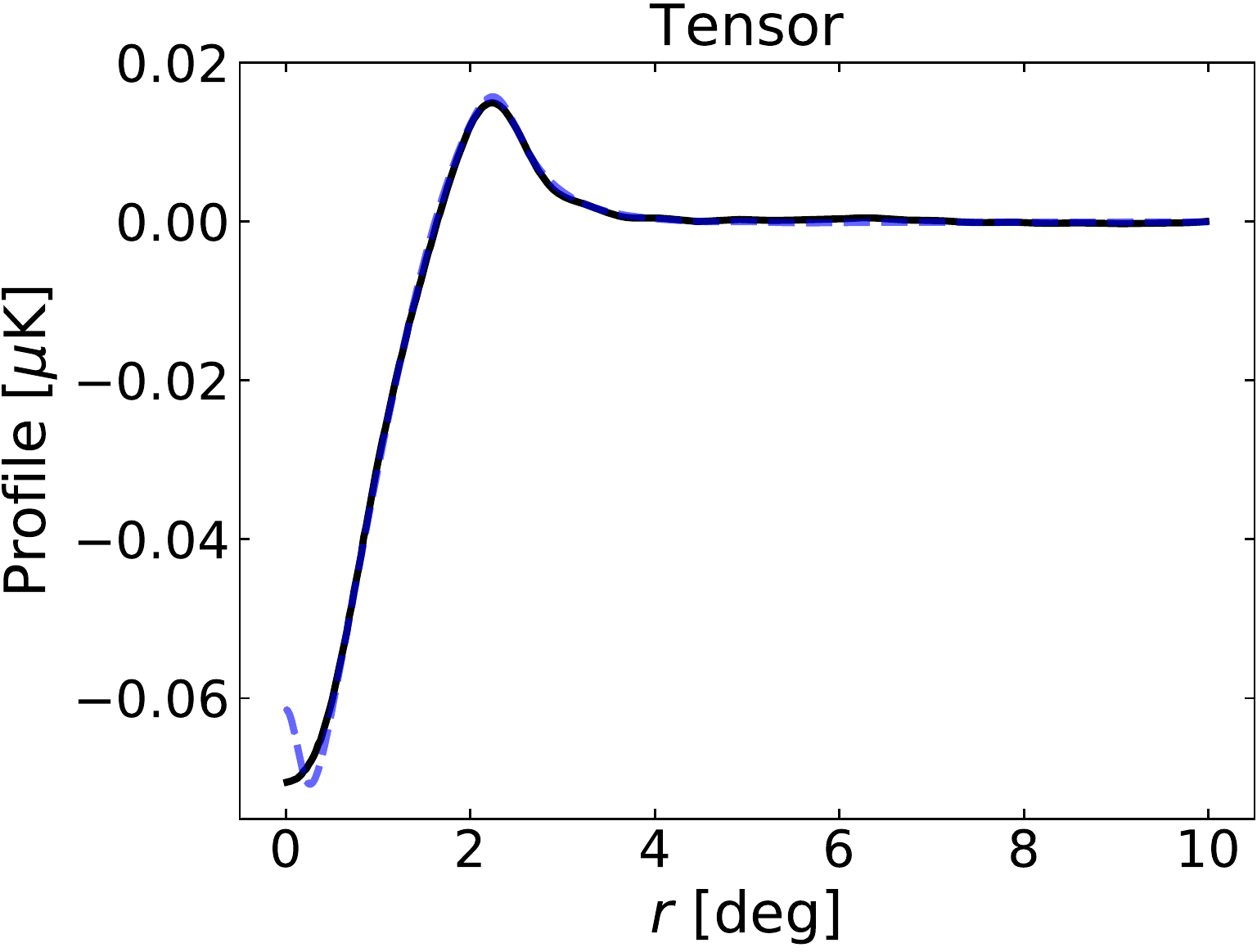}
\caption{Radial profiles computed from the azimuthal average of the $m=0$ modes for the $Q$ stacks on temperature saddle points for the simulated scalar, vector, and tensor maps, in black. The dashed blue lines are the theoretically predicted profiles in each case, obtained from eqs.~\ref{eq:TEBk}--\ref{eq:QrUrprof}, as described in Section~\ref{sec:saddle_stats}. Note that eq.~\ref{eq:QrUrprof} yields the profiles for $Q_\mathrm{r}$ and $U_\mathrm{r}$, as opposed to $Q$ and $U$. Since $Q_\mathrm{r}$ and $U_\mathrm{r}$ are undefined at the origin, this yields the observed departure between the stacks and predicted profiles as $r \to 0$.}
\label{fig:Q0_rad}
\end{figure}

We can separate out this monopole component of $Q$ to analyse it separately from the $m=4$ mode. Figure~\ref{fig:Q4_U} shows the $m=4$ component of $Q$ compared with the full $U$ stack for the scalar maps. We find that $Q_{m=4}$ is simply a $22.5^\circ$ rotation of $U$. That this should be the case becomes apparent when we consider the fact that the Fourier mode $(Q+iU)_{m=4} = Q_{m=4} + iU$ (since $U$ only has an $m=4$ mode to begin with). Thus $Q_{m=4}$ and $U$ are simply the real and imaginary parts, respectively, of a single $m=4$ Fourier mode. In other words, $Q_{m=4}$ is modulated by $\cos(4\phi)$ and $U_{m=4}$ is modulated by $\sin(4\phi)$, and are, therefore, simply $\frac{\pi}{8}$ ($22.5^\circ$) rotations of each other. The stacked polarization is, then, fully specified by $U$ and the monopole contribution to $Q$. Since $m=0$ modes are completely radially symmetric, it is straightforward to compute the radial profile from the azimuthal average of $Q_{m=0}$. We do this for the simulated scalar, vector, and tensor maps and show the results in Figure~\ref{fig:Q0_rad}. As demonstrated in Section~\ref{sec:saddle_stats}, the profile of the polarization stacked on temperature saddle points is determined by the two-point correlation functions, i.e. the $TE$ and $TB$ power spectra. From this, the features of the radial profiles in Figure~\ref{fig:Q0_rad} can be qualitatively understood in a similar way to the extrema stacks, as described in section 8 of Ref.~\citep{IandS2015}.

\begin{figure}[htb]
\centering
\includegraphics[width=0.8\columnwidth]{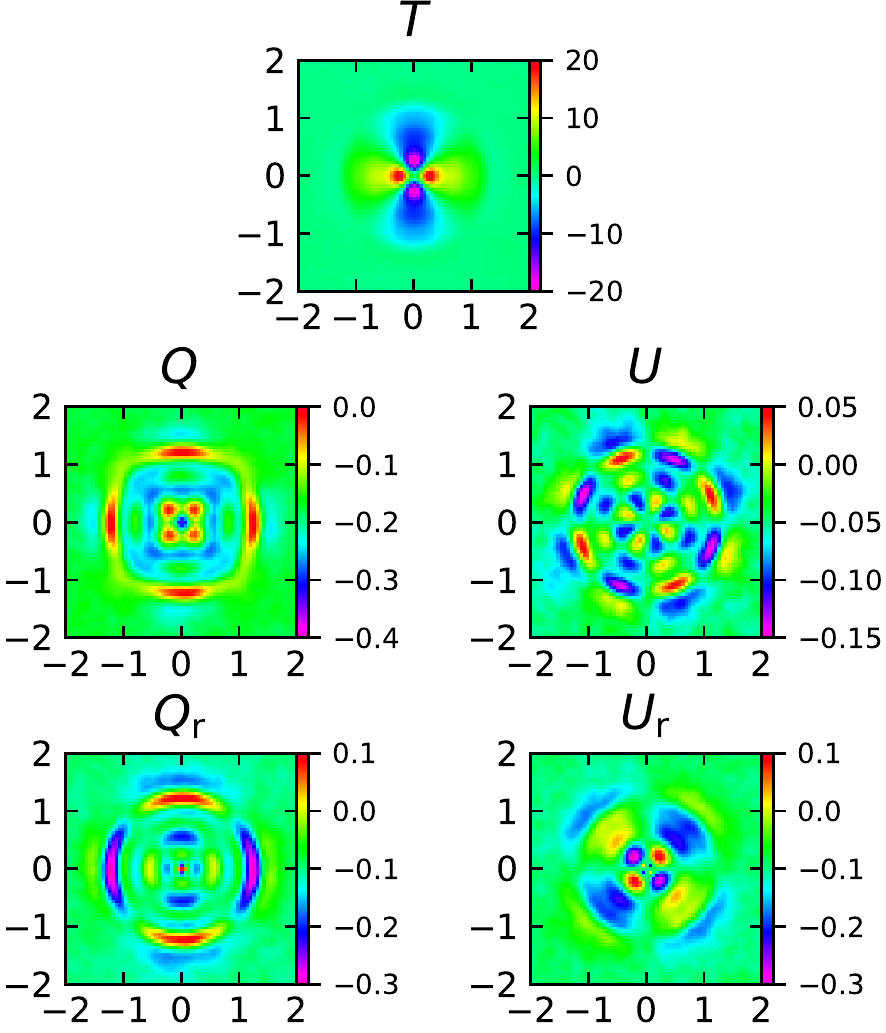}
\caption{Oriented stacks of $T$, $Q$, $U$, $Q_\mathrm{r}$, and $U_\mathrm{r}$ around temperature saddle points for the \textit{Planck} \texttt{SMICA} CMB map. The stacks are in units of $\mu$K, and the $x$- and $y$-axes are in degrees. The stacks for the real data are visually hard to distinguish from the simulated scalar CMB stacks.}
\label{fig:saddle_SMICA}
\end{figure}

\begin{figure}[htb]
\centering
\includegraphics[width=0.8\columnwidth]{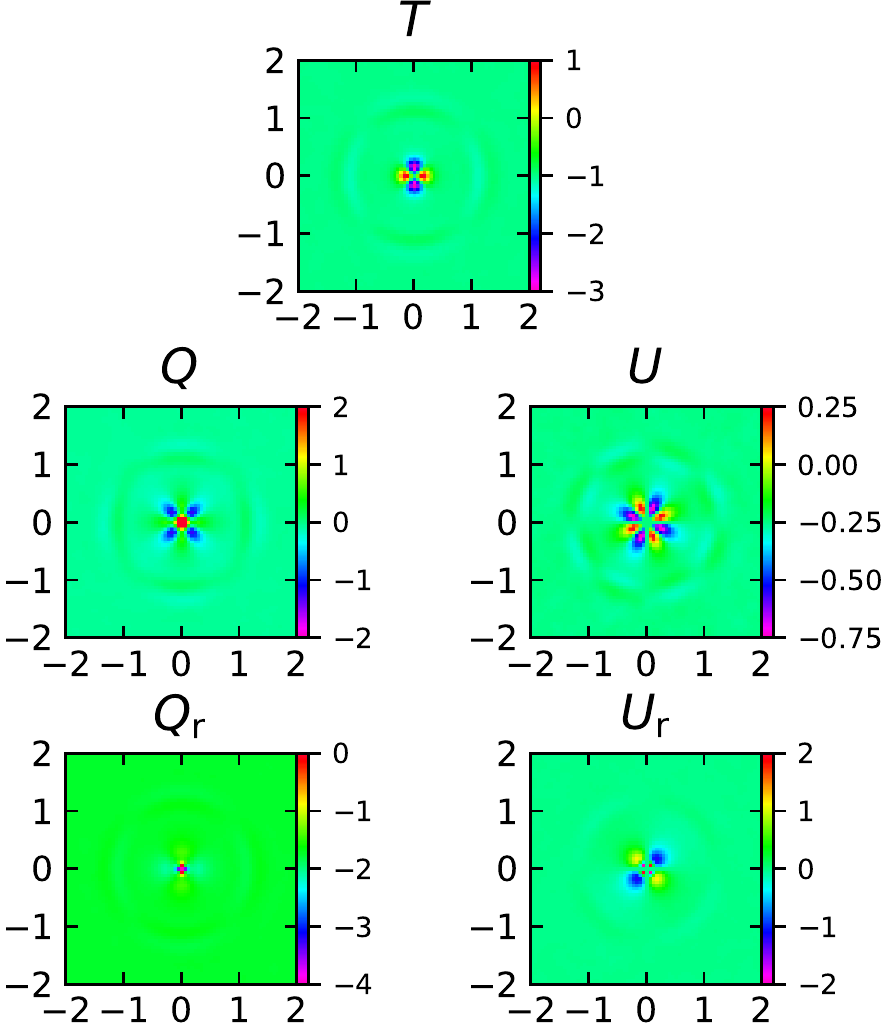}
\caption{Oriented stacks of $E$, $Q$, $U$, $Q_\mathrm{r}$, and $U_\mathrm{r}$ around $E$ saddle points for the \textit{Planck} \texttt{SMICA} CMB map. The stacks are in units of $\mu$K, and the $x$- and $y$-axes are in units of degrees. Note the similarity with the temperature stacks for the simulated vector map shown in Figure~\ref{fig:TQUQrUrxsvt}.}
\label{fig:Esaddle_SMICA}
\end{figure}

So far in this paper, we have only plotted stacks for simulated data. Figure~\ref{fig:saddle_SMICA} shows the results of temperature saddle-point stacks on real CMB data, namely the \textit{Planck} 2018 \texttt{SMICA} CMB maps. As expected, the results are consistent with the simulated scalar CMB stacks. We can also apply the same stacking procedure to other kinds of CMB maps. In particular, since the $E$-modes define a scalar map with the same Gaussian statistics, as with temperature, it is natural to also stack on saddle points in $E$. Following the above procedure, we stack $E$, $Q$, and $U$ on $E$ saddle points. Figure~\ref{fig:Esaddle_SMICA} shows the result of these stacks. Note the similarity of these results to the temperature stacks for the simulated vector map, with the only major difference being the scale of the features in the stacks, which is because of the different power spectrum. 

Here we have shown the results of stacks performed specifically on the \texttt{SMICA} component-separated map; however, we have performed the same procedure on other \textit{Planck} component-separated maps (\texttt{Commander}, \texttt{NILC}, and \texttt{SEVEM} \citep{2018arXiv180706208P}) and have found consistent results.

\section{Saddle-point statistics}
\label{sec:saddle_stats}

Now we turn to a mathematical analysis of what we have plotted in the previous section. The statistics of extremal points as a subset of critical points in a Gaussian random field have been well-characterized in the literature \citep[e.g.\,Refs.][]{BBKS, BondEfs1987,Komatsu2011,MC2016}. Here we generalize some of the results to saddle points in order to calculate the profiles of temperature and polarization patterns around saddle points in the CMB. Specifically, we consider critical points in the CMB temperature, $T( \theta, \phi)$, on the sky, but the following analysis holds for any scalar field on a sphere, such as $E$ or $B$ modes. 

\subsection{Critical points in the flat-sky approximation}
\label{sec:saddle_profs_Fourier}

Critical points are completely characterized by their derivatives up to and including second order. That is, the statistics of critical points are determined by a total of six degrees of freedom. In order to compute any statistical properties of critical points, we need to know the probability distribution function of these degrees of freedom. Ref.~\citep{MC2016} presents a general formalism for the statistics of critical points on the spherical sky; however, here we work in the flat-sky approximation for the sake of simplicity. 

For a field, $T(x,y)$, on the flat sky, the peak degrees of freedom are given by
\begin{equation}
\mathbf{X} =
\begin{pmatrix}
T \\ \partial_x T \\ \partial_y T \\ \partial^2_{xx} T \\ \partial^2_{yy} T \\ \partial^2_{xy} T
\end{pmatrix}.
\end{equation}
For a Gaussian random field, these degrees of freedom follow a multivariate Gaussian distribution \citep{BBKS,MC2016}. In the case of the CMB anisotropy maps, the means of all of these quantities have been subtracted out, so that all of the statistical information lies in the covariance matrix $\langle \mathbf{X} \mathbf{X}^\dag \rangle$. 

To compute the covariance matrix it is convenient to work in Fourier space, defining the Fourier transform $\widetilde{T}$ to be
\begin{equation}
T(\pmb{\theta}) = \int d^{2}\Bell \, \widetilde{T}(\Bell) e^{i \Bell \cdot \pmb{\theta}},
\end{equation}
where 
\begin{equation}
\langle \widetilde{T}(\Bell) \widetilde{T}^*(\Bell^\prime) \rangle = P_{TT}(\ell)\delta^2(\Bell - \Bell^\prime).
\end{equation}
Here, $\Bell = (\ell_x,\ell_y) = (\ell\cos{\psi},\ell\sin{\psi})$, where $\psi$ is the angle measured from the $x$-axis. The entries of the covariance matrix are then simply integrals over the power spectrum $P_{TT}(\ell)$ with different factors of $\ell$ in the integrand for the order of the derivative. For example, $\langle T, \partial^2_{xx} T \rangle = \int \ell^2_x P_{TT}(\ell)d^2\ell = \pi \int P_{TT}(\ell) \ell^3 d\ell$. The rest of the entries are similarly easy to compute, and we have
\begin{equation}
\langle \mathbf{X}\mathbf{X}^\dag \rangle = \sigma^2
 \begin{pmatrix}
 1 & 0 & 0 & q^{-2} & q^{-2} & 0 \\
 0 & q^{-2}  & 0 & 0 & 0 & 0 \\
 0 & 0 & q^{-2} & 0 & 0 & 0 \\
 q^{-2} & 0 & 0 & 3a^{-4} & a^{-4} & 0 \\
 q^{-2} & 0 & 0 & a^{-4} & 3a^{-4} & 0 \\
 0 & 0 & 0 & 0 & 0 & a^{-4}
 \end{pmatrix},
\label{eq:XX}
\end{equation}
where
\begin{subequations}
\label{eq:sigma_a}
\begin{equation}
\sigma^2 \equiv 2\pi \int d\ell\,P_{TT}(\ell) \ell,
\end{equation}
\begin{equation}
\frac{\sigma^2}{q^2} \equiv \pi \int d\ell \, P_{TT}(\ell) \ell^3,
\end{equation}
\begin{equation}
\frac{\sigma^2}{a^4} \equiv \frac{\pi}{4} \int d\ell \, P_{TT}(\ell) \ell^5.
\end{equation}
\end{subequations}
With this covariance matrix we can, in principle, compute any statistical property of critical points. The statistics of saddle points can then be obtained by imposing the constraint $\det H < 0$.

\subsection{Saddle-point profiles}
\label{sec:saddle-point-profiles}

Here we compute the average profiles of $T$, $E$, and $B$ around temperature saddle points. We will work in Fourier space, and as with temperature, we define the Fourier transforms and power spectra of $E$ and $B$ to be
\begin{subequations}
\label{eq:FT_TE}
\begin{equation}
E(\pmb{\theta}) = \int d^{2}\Bell \, \widetilde{E}(\Bell) e^{i \Bell \cdot \pmb{\theta}},
\end{equation}
\begin{equation}
B(\pmb{\theta}) = \int d^{2}\Bell \, \widetilde{B}(\Bell) e^{i \Bell \cdot \pmb{\theta}},
\end{equation}
\end{subequations}
where
\begin{subequations}
\begin{equation}
\langle \widetilde{E}(\Bell) \widetilde{E}^*(\Bell^\prime) \rangle = P_{EE}(\ell)\delta^2(\Bell - \Bell^\prime),
\end{equation}
\begin{equation}
\langle \widetilde{B}(\Bell) \widetilde{B}^*(\Bell^\prime) \rangle = P_{BB}(\ell)\delta^2(\Bell - \Bell^\prime),
\end{equation}
\begin{equation}
\langle \widetilde{T}(\Bell) \widetilde{E}^*(\Bell^\prime) \rangle = P_{TE}(\ell)\delta^2(\Bell - \Bell^\prime),
\end{equation}
\begin{equation}
\langle \widetilde{T}(\Bell) \widetilde{B}^*(\Bell^\prime) \rangle = P_{TB}(\ell)\delta^2(\Bell - \Bell^\prime).
\end{equation}
\end{subequations}
We consider the case when $T$ has a saddle point in a small neighbourhood of the origin. As we explain in Appendix B, it is necessary to phrase the condition in this way, rather than imposing the condition that there be a saddle point
\textit{at} the origin, because the latter has probability zero and leads to conditional probabilities that are undefined. We assume that the saddle point is oriented with positive eigenvector along the $x$-axis, so that the Hessian has eigenvalues $\lambda_+ > 0$ and $\lambda_- < 0$. Thus, we have a set of constraints
\begin{equation}
\mathbf{g} \equiv
\begin{pmatrix}
\partial_x T \\
\partial_y T \\
\partial^2_{xx} T \\
\partial^2_{yy} T \\
\partial^2_{xy} T
\end{pmatrix} = 
\begin{pmatrix}
0 \\
0 \\
\lambda_+ \\
\lambda_- \\
0
\end{pmatrix}.
\label{eq:g}
\end{equation}
Since we want to average all saddle points, we will eventually want to marginalize over all $\lambda_+$ and $\lambda_-$ values.

We want to know the mean values of $T$, $E$, and $B$ subject to these constraints. In Fourier space, we are interested in 
\begin{equation}
\overline{T}(\Bell) \equiv \langle \widetilde{T}(\Bell) | \mathbf{g} \rangle,
\label{eq:Tbar}
\end{equation}
and similarly for $\overline{E}$ and $\overline{B}$. That is, we want the conditional expectation of $T$, $E$, and $B$ given $\mathbf{g}$.

All of the quantities we are interested in are drawn from multivariate Gaussian distributions. In general, for a Gaussian random vector $\mathbf{h}$, subject to constraint $\mathbf{g}$, the expectation is given by
\begin{equation}
\langle \mathbf{h} | \mathbf{g} \rangle = \langle \mathbf{h} \mathbf{g}^{\dag} \rangle \langle \mathbf{g}\mathbf{g}^\dag \rangle ^{-1} \mathbf{g}.
\end{equation}
In our case, $\mathbf{h}$ consists of the Fourier modes $\widetilde{T}(\Bell)$, $\widetilde{E}(\Bell)$, $\widetilde{B}(\Bell)$.

Thus, we need to compute the covariance matrix $\langle \mathbf{g}\mathbf{g}^\dag \rangle$, but this is just the covariance matrix $\langle \mathbf{X X}^\dag \rangle$, given by eq.~\ref{eq:XX}, restricted to the lower-right $5 \times 5$ sub-matrix. It follows that
\begin{equation}
\langle \mathbf{g}\mathbf{g}^\dag \rangle^{-1} \mathbf{g} =
\begin{pmatrix}
0 \\ 0 \\ \frac{a^4}{8\sigma^2}(3\lambda_+ - \lambda_-) \\ \frac{a^4}{8\sigma^2}(3\lambda_- - \lambda_+) \\ 0
\end{pmatrix}
\equiv
\begin{pmatrix}
0 \\ 0 \\ \alpha_+ \\ \alpha_- \\ 0
\end{pmatrix}.
\label{eq:ggTg}
\end{equation}

Now, to obtain the expectation of any element of $\mathbf{h}$, we need to multiply the corresponding row of $\langle \mathbf{h} \mathbf{g}^\dag \rangle$ by the vector expressed in eq.~\ref{eq:ggTg}. This gives us
\begin{equation}
\overline{Z}(\Bell) = \alpha_+ \langle \widetilde{Z}(\Bell) \partial^2T_{xx}(0) \rangle + \alpha_- \langle \widetilde{Z}(\Bell) \partial^2T_{yy}(0) \rangle,
\label{eq:Zbar}
\end{equation} 
where $Z$ is $T$, $E$, or $B$. We can easily compute these last expectations, since, for instance,
\begin{equation}
\partial^2_{xx}T(0) = -\int d^2\Bell^\prime \widetilde{T}^*(\Bell^\prime) \ell^{\prime 2}_x,
\end{equation}
so
\begin{equation}
 \langle \widetilde{T}(\Bell) \partial^2T_{xx}(0) \rangle = - \int d^2 \Bell^\prime \ell^{\prime 2}_x \langle \widetilde{T}(\Bell) \widetilde{T}^* (\Bell^\prime) \rangle = -l^2_x P_{TT}(\ell).
\end{equation}
Thus, in the end we obtain
\begin{subequations}
\begin{equation}
\overline{T}(\Bell) = -(\alpha_+ \ell^2_x + \alpha_- \ell^2_y) P_{TT} (\ell),
\end{equation}
\begin{equation}
\overline{E}(\Bell) = -(\alpha_+ \ell^2_x + \alpha_- \ell^2_y) P_{TE} (\ell).
\end{equation}
\begin{equation}
\overline{B}(\Bell) = -(\alpha_+ \ell^2_x + \alpha_- \ell^2_y) P_{TB} (\ell).
\end{equation}
\end{subequations}

We now want to marginalize over $\lambda_+$ and $\lambda_-$. Since the results are linear in these quantities, this amounts to simply replacing them with their average values. Appendix~\ref{sec:math_thing} contains a calculation of these means, leading to the result $\lambda_\pm = \pm1.027\sigma/a^2$. Using this, we obtain
\begin{subequations}
\begin{equation}
\overline{T}(\Bell) = -0.513\frac{a^2}{\sigma}(\ell^2_x - \ell^2_y) P_{TT} (\ell),
\end{equation}
\begin{equation}
\overline{E}(\Bell) = -0.513\frac{a^2}{\sigma}(\ell^2_x - \ell^2_y) P_{TE} (\ell),
\label{eq:Ek}
\end{equation}
\begin{equation}
\overline{B}(\Bell) = -0.513\frac{a^2}{\sigma}(\ell^2_x - \ell^2_y) P_{TB} (\ell).
\label{eq:Bk}
\end{equation}
\label{eq:TEBk}
\end{subequations}
This fully specifies the temperature and polarization profiles around temperature saddle points. As anticipated, the result gives the form of purely quadrupolar patterns in the temperature and polarization. We can then obtain the Stokes parameters using \citep{Komatsu2011}
\begin{subequations}
\begin{equation}
Q_\mathrm{r}(\mathbf{\theta}) = \int \frac{d^2 \Bell}{(2\pi)^2} \big\{\widetilde{E}(\Bell) \cos[2(\phi-\psi)] + \widetilde{B}(\Bell) \sin[2(\phi-\psi)]\big\}e^{i \Bell \cdot \mathbf{\theta}},
\end{equation}
\begin{equation}
U_\mathrm{r}(\mathbf{\theta}) = \int \frac{d^2 \Bell}{(2\pi)^2} \big\{\widetilde{E}(\Bell) \sin[2(\phi-\psi)] - \widetilde{B}(\Bell) \cos[2(\phi-\psi)]\big\}e^{i \Bell \cdot \mathbf{\theta}}.
\end{equation}
\label{eq:QrUrprof}
\end{subequations}
The same calculations can also be done for the profiles of saddle points in $E$. Since $E$ and $T$ are both scalar Gaussian random fields, it is simply a matter of replacing every occurrence of ``$T$\,'' in the results for temperature saddle points with ``$E$\,''. 

Note that we have derived the expected profiles conditioning on there being a saddle point within some small neighbourhood of the origin. We could have done the calculation with the condition that the critical point be located exactly at the origin -- indeed, this may seem like the most natural thing to do. However, such a calculation would lead to a different numerical prefactor in eqs.~\ref{eq:TEBk}. We compare the two approaches explicitly in Appendix B. We show that conditioning on having a critical point at the origin is incorrect, because it involves conditioning on an event of probability zero. The latter approach leads to mathematical inconsistencies. In particular, it implies that
$1/\sqrt{3}$ of all critical points are saddle points, contradicting a theorem from Morse theory that
says that the ratio should be approximately $1/2$.

We note that the latter approach leads to mathematical inconsistencies. For example, one can naively compute the number of extrema and saddle points by conditioning on being at a critical point, and then compute the relative probabilities of that point being an extremum or a saddle point. This calculation gives the result that saddle points make up $1/\sqrt{3} \approx 58\%$ of all critical points, whereas a well-known theorem from Morse theory, and effectively going back to Euler and Poincar\'{e}, states that for analytic functions in the plane the numbers of saddle points and extrema are equal, and that they differ by exactly 2 points on the sphere \citep{Arango2006}.
 
\begin{figure}[htb]
\centering
\includegraphics[width=\columnwidth]{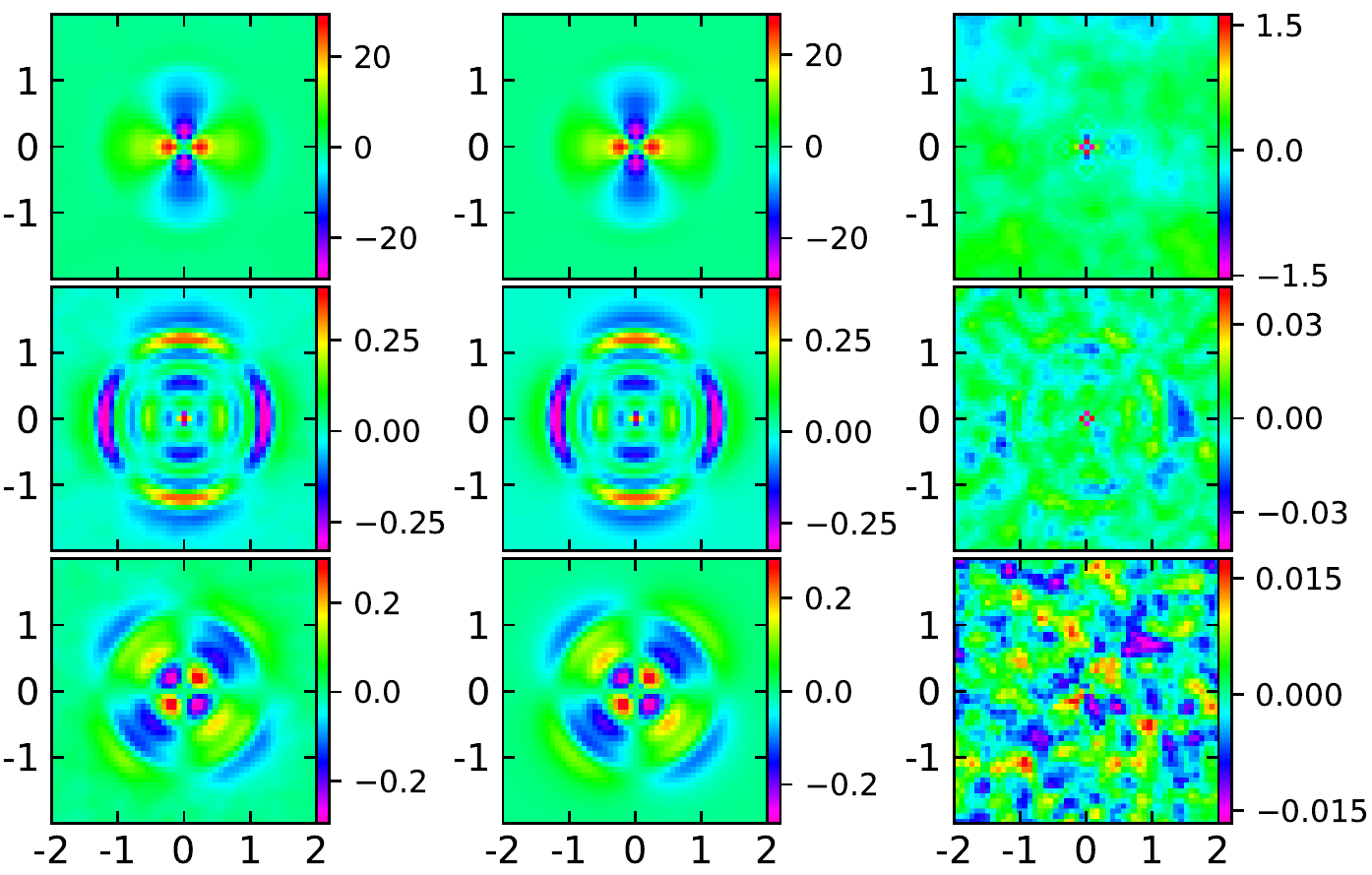}
\caption{Comparison between the $T$, $Q_\mathrm{r}$, and $U_\mathrm{r}$ oriented stacks on temperature saddle points for the simulated scalar map (left column), and their profiles computed according to the model expressed by eq.~\ref{eq:TEBk} (middle column). The residuals were then computed (right column). Note that the residual panels have a different contrast range than the signal panels. We carried out the same procedure for the vector and tensor maps and found similar consistency.}
\label{fig:model_resid}
\end{figure}

\subsection{Saddle point profiles in the curved sky}
\label{sec:cp_params}

The previous calculation gives the saddle-point profiles in the flat-sky approximation, which is more than sufficient here, since we are considering stacks for at most $20^\circ \times 20^\circ$ patches of the sky. However, Ref.~\citep{MC2016} presents a general formalism for critical point statistics that can be used to compute the saddle-point profiles on the sphere, which we will briefly describe here for completeness. In this formalism, the Hessian is decomposed into two parts, namely a part with non-vanishing trace and a traceless part. That is, in a local basis around the critical point, where the basis vectors $\mathbf{e}_x$ and $\mathbf{e}_y$ correspond to the usual spherical coordinate basis vectors $\bf{e}_\theta$ and $\bf{e}_\phi$, we can write the Hessian as
\begin{equation}
H[T] = 
\begin{bmatrix}
\partial^2_x T & \partial_x \partial_y T \\
\partial_x \partial_y T & \partial^2_y T
\end{bmatrix}
=
\frac{1}{2}
    \begin{bmatrix}
    \nabla^2 T & 0 \\
    0 & \nabla^2 T
    \end{bmatrix}
+ \frac{1}{2}
    \begin{bmatrix}
    \operatorname{Re}(\slashed{\partial}^{*})^2 T & -\operatorname{Im}(\slashed{\partial}^{*})^2 T \\
    -\operatorname{Im}(\slashed{\partial}^{*})^2 T & -\operatorname{Re}(\slashed{\partial}^{*})^2 T
    \end{bmatrix}.
\end{equation}
Here the derivatives $\slashed{\partial}$ and $\slashed{\partial}^*$ are the spin ladder operators on the sphere, which are proportional to the covariant derivatives on the sphere in the helicity basis (i.e. $\mathbf{e}_\pm = (\mathbf{e}_\theta \pm \mathbf{e}_\phi)/  \sqrt{2}$). In the local, flat coordinates the ladder operator can be written as $\slashed{\partial}^* = -\partial_x +i\partial_y$. The Hessian of a critical point contains information about the curvature around that point. In separating out the traceless part of the Hessian, we have decomposed the Hessian into one part describing the magnitude of the curvature, characterized by the Laplacian $\nabla^2 T$, and another describing the eccentricity of the curvature, characterized by the derivative $(\slashed{\partial}^*)^2 T$.

In this way, we can define four critical-point parameters, corresponding to the value of the field at the point, the first derivatives, and the magnitude and eccentricity of the local curvature. That is, one can define \citep{MC2016}
\begin{equation}
\begin{aligned}
& \nu \equiv \frac{T}{\sigma_\nu}, & \kappa \equiv -\frac{\nabla^2 T}{\sigma_\kappa}, \\
& \eta \equiv \frac{\slashed{\partial}^{*} T}{\sigma_\eta}, & \epsilon \equiv \frac{(\slashed{\partial}^*)^2 T}{\sigma_\epsilon},
\end{aligned}
\end{equation}
where we have divided by the square roots of the variances, $\sigma^2_\nu = \langle T^2 \rangle$, $\sigma^2_\eta = \langle |\slashed{\partial}^* T|^2 \rangle$, $\sigma^2_\kappa = \langle | \nabla^2 T|^2 \rangle$, and $\sigma^2_\epsilon = \langle |(\slashed{\partial}^*)^2 T|^2 \rangle$, so that the parameters are normalized to unit variance. Since $\eta$ and $\epsilon$ are complex numbers, there are once again six degrees of freedom characterizing the statistics of critical points. For a Gaussian random field, the probability distribution over the variables $\nu$, $\kappa$, $\eta$ and $\epsilon$ is Gaussian, with covariance matrix \citep{BBKS,MC2016}
\begin{equation}
S = 
\begin{pmatrix}
1 & \rho & 0 & 0 \\
\rho & 1 & 0 & 0 \\
0 & 0 & 1 & 0 \\
0 & 0 & 0 & 1
\end{pmatrix},
\end{equation} 
where $\rho = \sigma^2_\eta/\sigma_\nu \sigma_\kappa$.  

In order to compute the saddle-point profiles, we can expand fields on the sphere according to
\begin{subequations}
\begin{align}
f(\theta, \phi) &= \sum_{m=-\infty}^{\infty} f_m(\theta) e^{im\phi}, \\
f_m(\theta) &= \frac{1}{2\pi} \int d\phi f(\theta, \phi) e^{-im\phi}.
\end{align}
\end{subequations}
Note that if $f$ is a real field, then we have the property that $f_m = f_{-m}^*$. We then need to compute the expected radial profiles $\langle T_m(\theta) \rangle$, $\langle Q_{\mathrm{r},m}(\theta) \rangle$, and $\langle U_{\mathrm{r},m}(\theta) \rangle$, where $Q_\mathrm{r}$ and $U_\mathrm{r}$ are the radial Stokes parameters for a saddle point at $\theta=0$. Since critical points are completely characterized by their value, curvature, and eccentricity (together with the fact that they have vanishing first derivatives), the only non-vanishing parts are the $m=0$ and $m=2$ modes \citep{MC2016}, which depend on the spherical-harmonic power spectra $C^{TT}_\ell$, $C^{TE}_\ell$ and $C^{TB}$, and the constraints on the peak variables $\nu$, $\eta$, $\kappa$, and $\epsilon$. In particular, the critical-point profiles depend on the mean values of the peak parameters subject to these constraints \citep{MC2016}. Thus, in order to compute the expected saddle-point profiles, we need to determine the biases $b_\nu$, $b_\kappa$, and $b_\epsilon$, given by
\begin{subequations}
\begin{equation}
	\begin{pmatrix}
	b_\nu \sigma_\nu \\
	b_\kappa \sigma_\kappa
	\end{pmatrix}
	=
	\Sigma^{-1}
	\begin{pmatrix}
	\langle \nu \rangle \\
	\langle \kappa \rangle
	\end{pmatrix},
\label{eq:vk_bias}
\end{equation}
\begin{equation}
b_\epsilon = \frac{\langle \epsilon \rangle}{\sigma_\epsilon},
\end{equation}
\end{subequations}
where $\Sigma$ is the covariance matrix between $\nu$ and $\kappa$. We compute these expectations subject to the constraint that there is a critical point in a small neighbourhood around the origin, and that $|\epsilon | \geq \sqrt{a} |\kappa |$, where $a \equiv \sigma^2_\kappa / \sigma^2_\epsilon$, which selects for saddle points. We stress again that simply conditioning on $\eta = 0$ at the origin leads to inconsistent results. Instead, we impose the constraint that $\eta = 0$ for some point in a neighbourhood around the origin, which amounts to multiplying the integrand of the expectation integral by the determinant of the Hessian, which, in terms of the peak variables, is given by $\det(H) = (1/4) [\sigma^2_\kappa \kappa^2 - \sigma^2_\epsilon (\Re(\epsilon))^2 - \sigma^2_\epsilon (\Im(\epsilon))^2]$. Intuitively, we might expect to find that $b_\nu$ and $b_\kappa$ vanish for saddle points, since there is no reason for there to be a preference for saddle points to be either positive or negative in value or in curvature. We numerically verify that, indeed, $b_\nu$ and $b_\kappa$ are much less than $b_\epsilon$ for saddle points, so that only the $m=2$ mode effectively contributes to the saddle-point profiles.  Note, also, that in general the bias $b_\epsilon$ is a complex number; however, we can remove the imaginary part by rotating to the basis defined by the principle axes of the saddle point, since $\Im(\epsilon) \propto \partial_x \partial_y T$, which vanishes along the principle axes. In this paper, we always compute oriented saddle point stacks so that $b_\epsilon$ is real.

The $m=2$ radial profiles are computed to be \citep{MC2016}
\begin{subequations}
\label{eqn:x2}
\begin{equation}
\langle T_2(\theta) \rangle = b_\epsilon \sum_{\ell = 0}^\infty
\frac{2\ell+1}{4\pi} (W^T_\ell)^2 C_\ell^{TT} \ P_\ell^2(\cos \theta),
\label{eqn:t2}
\end{equation}
\begin{equation}
\langle Q_{r2}(\theta) \rangle = -2 b_\epsilon \sum_{\ell = 0}^\infty
\frac{2\ell+1}{4\pi} \sqrt{\frac{(\ell-2)!}{(\ell+2)!}} W^T_\ell W^P_\ell \left[ C_\ell^{TE}
\ P_\ell^{+}(\cos \theta) + i C_\ell^{TB} \ P_\ell^{-}(\cos \theta)
\right],
\label{eqn:qr2}
\end{equation}
\begin{equation}
\langle U_{r2}(\theta) \rangle = 2 i b_\epsilon \sum_{\ell = 0}^\infty
\frac{2\ell+1}{4\pi} \sqrt{\frac{(\ell-2)!}{(\ell+2)!}} W^T_\ell W^P_\ell \left[ C_\ell^{TE}
\ P_\ell^{-}(\cos \theta) + i C_\ell^{TB} \ P_\ell^{+}(\cos \theta)
\right].
\label{eqn:ur2}
\end{equation}
\label{eqn:sdlprofs}
\end{subequations}Here, the $P^m_\ell$ are the associated Legendre polynomials, and $W^T_\ell$ and $W^P_\ell$ are the spherical harmonic transformations of the smoothing functions (i.e.\,the beam) applied to the temperature and polarization maps. We have also written
\begin{equation}
\begin{aligned}
P_\ell^{+}(x) &= - \left[ \frac{\ell-4}{1-x^2} + \frac{1}{2} \ell
  \left(\ell-1\right) \right] P_\ell^2(x) + \left( \ell + 2 \right)
\frac{x}{1-x^2} P_{\ell-1}^2(x), \\
P_\ell^{-}(x) &= - 2 \left[ \left( \ell - 1 \right) \frac{x}{1-x^2}
  P_\ell^2(x) - \left( \ell + 2 \right) \frac{1}{1-x^2}
  P_{\ell-1}^2(x) \right].
\label{eqn:pl+_pl-}
\end{aligned}
\end{equation}
Thus, the saddle-point profiles depend only on the eccentricity bias, and the power spectra $C^{TT}_\ell$ and $C^{TE}_\ell$, since, in general, we have $C^{TB}_\ell = 0$. Equation~\ref{eqn:sdlprofs} can be shown to reduce to eq.~\ref{eq:TEBk} in the flat-sky limit (i.e.\,for $\ell \gg 1$, $\theta \ll 1$, and $\ell\theta \sim 1$) using the fact that $\epsilon = (\lambda_+ - \lambda_-)/\sigma_\epsilon$, $\int_\alpha^{2\pi+\alpha} d\psi \cos[2(\phi-\psi)]e^{ix\cos(\phi-\psi)} = -2\pi J_2 (l\theta)$, and the substitutions for going to the flat-sky approximation given in Appendix C of Ref.\,\citep{MC2016}.

\section{Constraints on cosmic birefringence}
\label{sec:biref}

Equipped with the theoretical calculations for the saddle-point profiles, as well as the extremum profiles given in Refs.\,\citep{Komatsu2011} and \citep{MC2016}, we can, in principle, use stacks of the data to constrain any model that affects the CMB power spectra, particularly models that alter the polarization spectra. It is clear, in the case of extrema, that when using hot spots one can add the constraining power of cold spots as independent information \citep{Komatsu2011}.  However, two obvious questions arise when extending this to saddle points: how do the constraints from saddle-point stacks compare to hot- and cold-spot stacks? and to what extent are the saddle-point constraints independent? Here we use saddle-point stacks to test cosmic birefringence as an example of their utility in constraining physical models. We compare the performance of saddle-point stacks with a similar test using temperature hot and cold spots previously performed in Ref.\,\citep{2016A&A...596A.110P}. We stress that, for the purposes of this paper, birefringence is just chosen as an explicit example of physics that could be constrained using stacks. We expect the lessons learned in this section to apply generally.

Models of parity violation in the electromagnetic sector of the Standard Model of particle physics predict a rotation of the plane of polarization of photons as they travel in vacuum \citep{PhysRevD.41.1231}. Within such models, the polarization of photons emitted from the last-scattering surface will rotate as the photons travel through the Universe towards our detectors. The amount of rotation, $\alpha$, is the cosmic birefringence angle, which can, in principle, vary with direction \citep{2017JCAP...12..046C}. Here, we only consider the case in which $\alpha$ is the same in all directions. The effect of a non-zero $\alpha$ results in a mixing of the $E$- and $B$-modes of the CMB polarization, in addition to the generation of $TB$ and $EB$ correlations that would otherwise vanish. Explicitly, for a constant $\alpha$, we can write the observed power spectra as \citep{PhysRevD.41.1231}
\begin{subequations}
\begin{align}
C^{\prime TT}_\ell &= C^{TT}_\ell, \\
C^{\prime EE}_\ell &= C^{EE}_\ell \cos^2(2\alpha) + C^{BB}_\ell \sin^2 (2\alpha), \\
C^{\prime BB}_\ell &= C^{EE}_\ell \sin^2 (2\alpha) + C^{BB}_\ell \cos^2 (2\alpha), \\
C^{\prime TE}_\ell &= C^{TE}_\ell \cos(2\alpha), \\
C^{\prime TB}_\ell &= C^{TE}_\ell \sin(2\alpha), \\
C^{\prime {EB}}_\ell &= \frac{1}{2} (C^{EE}_\ell - C^{BB}_\ell) \sin(4\alpha),
\end{align}
\label{eqn:biref_spec}
\end{subequations}
Here, the $C^\prime_\ell$ are the observed spectra, while the unprimed $C_\ell$s are the spectra for a universe without birefringence. The modulation of the $TE$ correlation and generation of the $TB$ correlation naturally results in different profiles for the polarization stacks around temperature saddle-points compared to the case of $\alpha = 0$, as seen clearly in eq.~\ref{eqn:sdlprofs}. Thus, we can potentially use the saddle-point stacks to compute likelihoods for constraining the value of $\alpha$. 

In order to test the constraining power of the saddle-point stacks compared to the extrema stacks, we generate birefringent maps with a known $\alpha$ from the simulated scalar power spectra generated by \texttt{CAMB}. It is simple to generate birefringent maps, since the new polarization is simply a rotation of the old polarization by $\alpha$, which can be computed from eq.~\ref{eqn:polrot}. An estimate of the signal-to-noise ratio, $S/N$, for the angle $\alpha$ can be obtained from the power spectra. For small $\alpha$, and in the case where $C^{BB}_\ell = 0$, the signal manifests predominantly in the generation of $TB$ correlations (as well as $EB$ correlations; however, for simplicity we restrict our attention to $TB$ in the following analysis). Then the signal-to-noise is just the signal-to-noise of the $TB$ correlations, which is given by \citep{2016JCAP...06..046S}
\begin{equation}
\frac{S}{N} = \sum_{\ell=2}^{\infty} \frac{(2\ell + 1) C^{\prime TB}_\ell}{ \widetilde{C}^{BB}_\ell \widetilde{C}^{TT}_\ell  + (\widetilde{C}^{TB}_\ell)^2},
\label{eqn:clSN}
\end{equation}
where $\widetilde{C_\ell}$ indicates the power spectra in the case of no birefringence, but with noise. Note that if the birefringent maps are generated using power spectra without noise, then the signal-to-noise ratio is infinite in the case of vanishing $BB$ correlations. For current experiments, the signal-to-noise is most sensitive to noise in the polarization, and remains relatively unaffected by noise in the temperature. Therefore we choose to add noise only to the polarization. The noisy power spectra are then given by
\begin{subequations}
\begin{align}
\widetilde{C}^{EE}_\ell &= C^{EE}_\ell + \frac{n^2}{(W^P_\ell)^2}, \\
\widetilde{C}^{BB}_\ell &= C^{BB}_\ell + \frac{n^2}{(W^P_\ell)^2}, 
\end{align}
\end{subequations}
where $W^P_\ell$ is the polarization beam function, and $n$ is a constant parameterizing the amount of noise. Since we are interested in a comparison of the performance of the saddle-point and extrema stacks, the choice of $n$ is arbitrary. Here we choose $n$ = 0.04\,$\mu$K. 

In order to recover the value of $\alpha$ using saddle-point stacks, we need to compute the residuals from the stacks performed on the (simulated) data, and the predicted profile of the stacks computed from the power spectra. Using a uniform prior on $\alpha$ (for the range $( -\frac{\pi}{2} , \frac{\pi}{2} ]$), we have
\begin{equation}
P(\alpha | d) \propto P(d | \alpha),
\end{equation}
with
\begin{equation}
P(d | \alpha) = \frac{1}{\sqrt{2\pi |\Sigma |}} \exp{-\frac{1}{2} (d - (Q_\mathrm{r},U_\mathrm{r})(\alpha))^\mathsf{T} \Sigma^{-1} (d - (Q_\mathrm{r},U_\mathrm{r})(\alpha))}.
\end{equation}
Here, $d$ is the data, i.e. the stacked images of $Q_\mathrm{r}$ and $U_\mathrm{r}$, and  $(Q_\mathrm{r},U_\mathrm{r})(\alpha)$ are the theoretical profiles computed as a function of $\alpha$ given by eqs.~\ref{eq:TEBk}--\ref{eq:QrUrprof} combined with eq.~\ref{eqn:biref_spec}. The quantity $\Sigma$ is the covariance matrix of the saddle-point stacks and is computed explicitly for general stacks in Ref.\,\citep{MC2016}. Here, we follow Ref.\,\citep{PhysRevD.41.1231} and estimate the covariance matrix by weighting the rms of the noise in polarization with the inverse of the total number of pixels used in each re-gridded pixel in the stacked image. We assume that the covariance is diagonal in pixel space, which, for large enough pixel bins, is a reasonable approximation.

We identify critical points in the CMB maps as points in the map of $|\nabla T|^2$ that are smaller than their eight nearest-neighbour pixels, as discussed in Section~\ref{sec:stacks}. Saddle points are then critical points with $\det(H)<0$ and extrema are critical points with $\det(H)>0$. Hot spots are distinguished from cold spots by the sign of the Laplacian, $\nabla^2 T$. This method of identifying hot and cold spots differs from the method used, for example, in Refs.~\citep{IandS2015,PhysRevD.41.1231}, where hot spots are identified as pixels in the map that have a greater value than their nearest-neighbour pixels, and similarly for cold spots. We, however, choose to find the extrema in a manner that is consistent with our method of finding saddle points, so that, for example, we find an equal number of extrema and saddle points, consistent with Morse theory \citep{Arango2006}. Using the nearest neighbour-method for identifying extrema returns 180,000 extrema, whereas the critical point method returns 274,000 extrema, which is equal to the number of saddle points found. This is an approximately 50\% increase in the number of extrema identified. Since not all minima are, in fact, zeros of the gradient, the critical point method will generally include points that are not true extrema. It is reasonable to ask, then, whether these additional points contain any information or are simply adding noise. In the present example of testing birefringence, we performed the analysis with both the critical-point and nearest-neighbour methods, and found that both methods recovered the same signal-to-noise ratio. Thus, while it is difficult to argue that the additional points are comprised of mostly genuine extrema, both sets of points contain the same information. Here we choose to use the critical point method for consistency with the saddle-point analysis.

For both the extrema and saddle points, we consider all points weighted uniformly, without proof that this is an optimal weighting for either case. The posterior on $\alpha$ is approximately Gaussian, so we only need two numbers to describe the recovered signal, namely, the mean, $\mu$, and the standard deviation, $\sigma$. To compare the constraining power of the saddle points and the extrema, we compare the signal-to-noise ratio, $\mu/\sigma$, of both methods.

We find that for $\alpha = 0.2^\circ$, the extrema recover a signal of $\alpha = 0.190^\circ \pm 0.030^\circ$, while the saddle points recover $\alpha = 0.190^\circ \pm 0.023^\circ$. The signal-to-noise ratio is, therefore, (rounding the results) 6 for the extrema, and 8 for the saddle points. Thus, in this specific example, the saddle-point stacks contain more constraining power than the hot and cold spots combined. One might imagine that, ideally, one would be able to combine the information from the saddle points and extrema to obtain even higher signal-to-noise. However, it is not immediately clear how the saddle points and extrema stacks covary, and thus how to combine the two. In the best case scenario, the saddle points and extrema would simply be independent, so that the noise effectively scales as the inverse of the square of the total number of saddle points and extrema combined. To gain some insight into whether this is the case, we can compare the signal-to-noise ratios of the stacks with the signal-to-noise estimated from the power spectra, given by eq.~\ref{eqn:clSN}. For $\alpha = 0.2^\circ$, the power spectra give an estimated signal-to-noise of 9. Under the assumption of Gaussianity, the power spectra contain all of the physical information in the CMB. Thus, we can take the estimate of the signal-to-noise from the power spectrum to be essentially optimal. With this assumption, we can see that the saddle points return a nearly optimal constraint. Thus, it must be that the information in the extrema and saddle points is at least partly redundant, and cannot be naively combined.

We have also tested that we obtain the same general trends for other values of $\alpha$. For example, for $\alpha=0.5$, the signal-to-noise for extrema is 16, whereas it is 23 for all saddle points. The signal-to-noise ratio estimated from the power spectra is also 23, and so in this case the saddle points achieve equivalent constraining power as the power spectra. Saddle-point stacks, therefore, appear to perform better than stacks on extrema when constraining non-standard physics. It is important to note that this is not a rigorously proven mathematical result; we have only shown that the saddle points identified by the critical-point algorithm constrain birefringence close to optimally and contain more information than the extrema identified by either the critical-point of nearest-neighbour algorithm. Since we do not claim that any of these algorithms identify saddle points or extrema perfectly, we are unable to compare the theoretical constraining power for these sets of points. However, we have empirically shown that our saddle point analysis outperforms the extrema analysis commonly used in the literature. Of course, such stacks do not outperform power-spectrum analysis (at least for Gaussian statistics), but there are occasions where there may be advantages to using a real-space approach. For one thing, stacks may be less susceptible to some kinds of systematic effects, and hence stacks (particularly on saddle points) provide a valuable alternative approach for testing any physics that might alter the power spectra.

\section{Conclusions}
\label{sec:conclusions}

Although stacks around hot spots and cold spots in CMB maps have been extensively studied, little attention has been given to saddle points. We have performed stacks of saddle points on simulated microwave skies, generated from purely scalar, vector, and tensor perturbations, as well as for real \textit{Planck} data. The features in these stacks are visually striking, and can also be used to characterize some of the physics underlying the CMB correlation functions.  Additionally, such saddle-point stacks can be used to constrain non-standard physics. To that end, we have computed the theoretically predicted profiles of the stacks, both for the spherical sky, and in the flat-sky approximation. As an example of using the stacks to test physics, we simulated birefringent skies and used saddle-point and extremal-point stacks to constrain the birefringent angle. Comparing the constraining power, we found that points identified as saddle points by the critical-point algorithm contain more information than points identified as extrema using either the critical-point or nearest-neighbour algorithms. Therefore, the addition of saddle points in tests of new physics, along with extremal points, has the potential to improve physical constraints. Indeed, the saddle-point stacks achieve a signal-to-noise ratio on the birefringence angle approaching the estimated signal-to-noise from the power spectrum, suggesting that saddle-point stacks alone may provide a close-to-optimal estimator of the birefringence angle. We also performed preliminary tests comparing saddle points to extrema for constraining the presence of tensor modes in simulated CMB maps and found, again, that saddle points have more constraining power; however, further work should be done to characterize the statistics of using stacks to constrain tensor modes, and other changes in the physics of the power spectrum. Although we have not investigated a wide range of examples, and have not provided a strict proof of any general claim, it appears to be the case in practice that saddle-point stacks provide a more sensitive test than stacks on extrema. Whether or not this is always the case, we certainly recommend that future tests of physics using stacks should also include saddle-point stacks, in addition to extrema, as a potentially stronger constraint.

\acknowledgments

This work was supported by the Natural Sciences and Engineering Research
Council of Canada (NSERC), and is based on observations obtained with \textit{Planck} (\url{http://www.esa.int/Planck}), an ESA science mission with instruments and contributions directly funded by ESA Member States, NASA, and the CSA. Some of the results in this paper have been derived using the \texttt{HEALPix} package. We also thank A. Marcos-Caballero for useful discussion.

\bibliographystyle{JHEP}

\bibliography{biblio}

\appendix
\section{Un-oriented stacks on hot and cold spots}
\label{sec:peak_stacks}

For the sake of comparison with the saddle-point stacks, in Figure~\ref{fig:hsp_SMICA} we show un-oriented stacks of temperature and polarization (given in terms of the Stokes parameters $Q$ and $U$) on temperature maxima (defined as pixels with temperature greater than the eight nearest neighbours) above a threshold $\nu=0$, for the \texttt{SMICA} data. The stacks are un-oriented in the sense that the patches around each hot spot are added together with random orientations with respect to one another. We also show the so-called radial Stokes parameters $Q_\mathrm{r}$ and $U_\mathrm{r}$, defined in Section~\ref{sec:ortstck}. We stress that the rotation of the Stokes parameters to obtain the radial Stokes parameters parameters is unrelated to the rotations performed to obtain oriented stacks. Stacks similar to those of Fig.~\ref{fig:hsp_SMICA} can be found elsewhere, for example, in Refs.~\citep{Komatsu2011} and \citep{IandS2015}.

In contrast to the stacks found in Refs.~\citep{Komatsu2011} and \citep{IandS2015}, which identify hot spots and cold spots as pixels whose value is greater or less than, respectively, their nearest-neighbour pixels, in the rest of this paper we first identify critical points as minima (with respect to their nearest neighbours) and then extrema as those critical points with positive $\det(H)$. Hot spots are, then, those pixels with $\nabla^2 T < 0$ and cold spots are those with $\nabla^2T > 0$. This method of finding extrema is consistent with our method of finding saddle points.

\section{The ratio of saddle points to critical points}
\label{sec:math_thing}

In this Appendix, we calculate the probability that a given 
critical point
is a saddle point. This appears to be a straightforward computation
of a conditional probability, but as we will see care must be taken in 
specifying the problem correctly. We work in the flat-sky
approximation throughout (but note that it is easy to extend the results to the full sky).

We will begin by deliberately performing an incorrect calculation, and then 
show how it must be corrected. We conclude with a calculation of the
mean Hessian eigenvalues for saddle points, which is needed in Section~\ref{sec:saddle-point-profiles}.

\subsection{Incorrect calculation.} Let $T$ be a realisation
of a Gaussian random process on the plane, following the notation of Section~\ref{sec:saddle_profs_Fourier}.
We calculate the conditional
probability that the origin is a saddle point, given that
it is a critical point.
The latter condition requires that $\partial_xT(0)=\partial_yT(0)=0$, while
the former requires that the determinant of the Hessian be negative.
We parameterize the Hessian matrix elements in terms of the quantities
\begin{align}
h_0&=\left(a^2\over\sigma\right)\left\{\partial_{xx}T(0)+\partial_{yy}T(0)\right\},\\
h_+&=\left(a^2\over\sigma\right)\left\{\partial_{xx}T(0)-\partial_{yy}T(0)\right\},\\
h_\times&=\left(a^2\over\sigma\right)2\partial_{xy}T(0).
\end{align}
Using the covariance matrix elements in eq.~\ref{eq:XX}, it is straightforward
to show that these are independent Gaussian random variables with
mean zero and variances 2, 1, and 1, respectively.

The determinant of the Hessian is 
\begin{equation}
{\rm det}({\bf H}) = {\sigma^2\over 4a^4}\left\{h_0^2-(h_+^2+h_\times^2)\right\},
\end{equation}
so the conditional probability we wish to compute is
\begin{equation}
p_{\rm saddle} \equiv
P\big(h_0^2<h_+^2+h_\times^2\, | \, \partial_xT(0)=\partial_yT(0)=0\big).
\label{eq:psaddle}
\end{equation}

The second derivatives of $T$ are all independent of the first 
derivatives, as seen in eq.~\ref{eq:XX} (or by a parity argument). So
we can ignore the condition in this expression and compute
the probability by integrating over the Gaussian distribution
of $h_0,h_+,h_\times$:
\begin{equation}
p_{\rm saddle} = \int_{h_0^2<h_+^2+h_\times^2}
G(h_0;0,2)G(h_+;0,1)G(h_\times;0,1)\,dh_0\,dh_+\,dh_\times
={1\over\sqrt{3}}.
\end{equation}
Here $G(x;\mu,\sigma^2)$ is the Gaussian distribution with mean $\mu$
and variance $\sigma^2$.

We conclude that $1/\sqrt{3}=0.577$ of all critical points are saddle points. This result is found in the literature, for example in Ref.~\citep{2003MNRAS.344..905D}, although it does not affect the main results of their paper. 

We can see that the above conclusion is incorrect by comparing with 
a well-known result from Morse theory, which says that all functions
on the sphere (subject to very mild hypotheses) have precisely
two more extrema than saddle points. In general, the number of saddle points exceeds the number of extrema in an analytic function on a manifold by the Euler characteristic of the manifold \citep{Arango2006}.
Suppose that we have a Gaussian
random process
on the sphere, whose power spectrum contains mostly small-scale power,
so that the flat-sky approximation is good. Then both the number
of extrema and the number of saddle points are large, and the fraction
of all critical points that are saddle points must be $1/2$, not
$1/\sqrt{3}$. 

\subsection{Correcting the problem.}
The error in the above calculation arises from the definition
of $p_{\rm saddle}$ in equation (\ref{eq:psaddle}). The
condition $\partial_xT(0)=\partial_yT(0)=0$ has probability zero, so 
the conditional probability is undefined. 

To solve this problem, we must replace that condition with the condition
$\partial_xT(\mathbf r)=\partial_yT(\mathbf r)=0$ for some $\mathbf r$ within
a given small neighbourhood ${\cal N}$ of the origin. In such a neighbourhood,
we can approximate
\begin{equation}
\nabla T(\boldsymbol{\theta}) = \nabla T(0)+{\bf H}\boldsymbol{\theta},
\end{equation}
where ${\bf H}$ is the Hessian matrix at the origin. There
will be a critical point in the neighbourhood as long
as $-\nabla T(0)$ lies within the neighbourhood ${\bf H}{\cal N}$.
The area of the latter is $|\mathrm{det}({\bf H})|$ times
the original area, and hence the probability of finding a critical
point in a neighbourhood, given ${\bf H}$, is also proportional to $|\mathrm{det}
({\bf H})|$. By Bayes's theorem, therefore, the 
probability density of $h_0,h_+,h_\times$, given that there
is a critical point in a small neighbourhood of the origin, is
\begin{equation}
f(\mathbf h \, | \, \mbox{nearby critical point})\propto f(\mathbf h) 
|\mathrm{det}({\bf H})|,
\label{eq:condprobh}
\end{equation}
where $f(\mathbf h)$ is the Gaussian distribution on $\mathbf h = (h_0,h_+,h_\times)$.

The fraction of all critical points that are saddle points then becomes
\begin{equation}
p_{\rm saddle} = {\int_{h_0^2<h_+^2+h_\times^2}
G(h_0;0,2)G(h_+;0,1)G(h_\times;0,1)|\mathrm{det}(\mathbf{H})|
\,dh_0\,dh_+\,dh_\times
\over 
\int
G(h_0;0,2)G(h_+;0,1)G(h_\times;0,1)|\mathrm{det}(\mathbf{H})|
\,dh_0\,dh_+\,dh_\times
}.
\end{equation} 
This expression evaluates to $1/2$, in agreement with Morse theory.

\subsection{Mean eigenvalues.}
In Section~\ref{sec:saddle_profs_Fourier}, we need the mean values of the Hessian eigenvalues, taken
over all saddle points. We can find this by integrating
over the probability density (\ref{eq:condprobh}).
The Hessian eigenvalues are
\begin{equation}
\lambda_\pm = {\sigma\over 2a^2}\left(h_0\pm\sqrt{h_+^2+h_\times^2}\right).
\end{equation}
So the mean over all saddle points is 
\begin{equation}
\bar\lambda_\pm = 
{\int_{h_0^2<h_+^2+h_\times^2}
\lambda_\pm G(h_0;0,2)G(h_+;0,1)G(h_\times;0,1)|\mathrm{det}(\mathbf{H})|
\,dh_0\,dh_+\,dh_\times
\over
\int_{h_0^2<h_+^2+h_\times^2}
G(h_0;0,2)G(h_+;0,1)G(h_\times;0,1)|\mathrm{det}(\mathbf{H})|\,dh_0\,dh_+\,dh_\times
},
\end{equation}
which evaluates to
\begin{equation}
\bar\lambda_\pm = \pm {\sigma\over a^2}\left(10+3\sqrt{2}\cot^{-1}\sqrt{2}
\over 4\sqrt{3\pi}\right)=1.027{\sigma\over a^2}.
\end{equation}

%

\end{document}